\documentclass[final,authoryear,11pt]{elsarticle}
\journal{Ocean Engineering}
\usepackage{fullpage}
\usepackage{endfloat}

\usepackage{natbib,upgreek}
\usepackage{mathtools}
\usepackage{rotating}
\usepackage{amsmath}
\usepackage{amssymb}
\usepackage{bbm}

\def\bSig\mathbf{\Sigma}

\newcommand{\un}[1]{\boldsymbol{#1}}
\newcommand{\pbi}{\begin{itemize}}
\newcommand{\pei}{\end{itemize}}

\newcommand{\pbc}{\begin{center}}
\newcommand{\pec}{\end{center}}
\newcommand{\pbe}{\begin{eqnarray*}}
\newcommand{\pee}{\end{eqnarray*}}
\newcommand{\pben}{\begin{eqnarray}}
\newcommand{\peen}{\end{eqnarray}}
\providecommand{\Pr}{\mathbb{Pr}} 

\providecommand{\xb}{\boldsymbol{x}}
\providecommand{\Xb}{\boldsymbol{X}}

\providecommand{\Ub}{\boldsymbol{U}}

\providecommand{\Xt}{{\tilde{X}}}

\providecommand{\Xtb}{\boldsymbol{\tilde{X}}}

\usepackage{tikz}
\usetikzlibrary{arrows}

\begin{document}

\setlength{\parindent}{0.5cm}

\begin{frontmatter}

\title{On environmental contours for marine and coastal design.}
\author[shellnl]{Emma Ross}
\author[dnvgl]{Ole Christian Astrup} 
\author[dnvgl]{Elzbieta Bitner-Gregersen}
\author[hrw]{Nigel Bunn}
\author[shellscot]{Graham Feld}
\author[hrw]{Ben Gouldby}
\author[uo]{Arne Huseby}
\author[hrw]{Ye Liu}
\author[shellnl]{David Randell}
\author[dnvgl]{Erik Vanem}
\author[shelluk,lancs]{Philip Jonathan\corref{cor1}}
\cortext[cor1]{Corresponding author. Email: {\tt philip.jonathan@shell.com}}
\address[shellnl]{Shell Global Solutions B.V., 1031 HW Amsterdam, The Netherlands.}
\address[dnvgl]{DNV-GL, 1363 Hovik, Norway.}
\address[hrw]{HR Wallingford, Wallingford OX10 8BA, United Kingdom.}
\address[shellscot]{Shell UK Ltd., Aberdeen AB12 3FY, United Kingdom.}
\address[uo]{Department of Mathematics, University of Oslo, 0316 Oslo, Norway.}
\address[shelluk]{Shell Research Ltd., London SE1 7NA, United Kingdom.}
\address[lancs]{Department of Mathematics and Statistics, Lancaster University, Lancaster LA1 4YW, United Kingdom.}
%

%
%

\begin{abstract}
Environmental contours are used in structural reliability analysis of marine and coastal structures as an approximate means to locate the boundary of the distribution of environmental variables, and hence sets of environmental conditions giving rise to extreme structural loads and responses. Outline guidance concerning the application of environmental contour methods is given in recent design guidelines from many organisations. However there is lack of clarity concerning the differences between approaches to environmental contour estimation reported in the literature, and regarding the relationship between the environmental contour, corresponding to some return period, and the extreme structural response for the same period. Hence there is uncertainty about precisely when environmental contours should be used, and how they should be used well. This article seeks to provide some assistance in understanding the fundamental issues regarding environmental contours and their use in structural reliability analysis.  Approaches to estimating the joint distribution of environmental variables, and to estimating environmental contours based on that distribution, are described. Simple software for estimation of the joint distribution, and hence environmental contours, is illustrated (and is freely available from the authors). Extra assumptions required to relate the characteristics of environmental contour to structural failure are outlined. Alternative response-based methods not requiring environmental contours are summarised. The results of an informal survey of the metocean user community regarding environmental contours are presented. Finally, recommendations about when and how environmental contour methods should be used are made.
\end{abstract}



\end{frontmatter}


\section{Introduction} \label{Sct:Int}
%
\subsection{Metocean design} \label{Sct:Int:MODsg}
%
Currently, numerous approaches for establishing design criteria for metocean loads and responses of marine structures and coastal facilities are used by different practitioners. Some of these are included in marine industry standards and guidelines; others are internal standards of different organisations active in ocean engineering. Rigorous comparison of some approaches has been reported in the literature, but there is still uncertainty in the user community regarding the relative merits of different approaches. Within the marine industry, estimation of a joint metocean description has been considered for more than thirty years. It was shown that typically, environmental forces on marine structures may be reduced by 5\% to 40\% by accounting for the lack of complete dependence between metocean variables (wind, wave, current, etc.) traditionally used in design (e.g. \citealt{EP85}, \citealt{FldEA18}). Development of reliability methods (e.g. \citealt{MdsEA86}) and their implementation by some parts of the industry in the 1980s brought joint probabilities into focus: they are required for a consistent treatment of the loading in Level III reliability analysis and for assessment of the relative importance of various metocean variables during extreme load and response conditions, fatigue damage and at failure.
 
Until the middle of the 1990s, very few metocean data sets of sufficient quality were available, limiting development of joint probability models: this has changed during the last twenty years. Comprehensive hindcast data are now available for locations world-wide, including simultaneous values for wind, waves, current, sea water level, ice and snow of sufficient quality and duration. Today joint probabilities are referenced in industry standards and guidelines (e.g. \citealt{IACS01}, \citealt{DNV-RP-C205:2017}, \citealt{NRS17}, \citealt{ISO19901-01:2015}).  They are required for application of the Formal Safety Assessment (FSA) methodology in rule development, providing risk based goal-oriented regulations that are well balanced with respect to acceptable risk levels and economic considerations, as recommended in \cite{IMO01}.

Different standards describing the application of joint probability methods exist. \cite{ISO19901-01:2015}, \cite{DNV-RP-C205:2017} and \cite{NRS17} suggest that joint probability methods should be applied if reliable simultaneous data exist. \cite{NRS17} further recommends that the duration of a data should be sufficiently long to capture the probability of occurrence for all combinations of importance regarding predictions of metocean actions and action effects. Further, in a case of wind, wave and current it recommends that at least three years of simultaneous data is required to characterise the lack of complete dependence between these variables reliably in design.

Given that it is possible to establish a model for the ``short-term'' distribution of response given sea state parameters (for example, for a three hour sea state), \cite{NRS17} states that the designer has essentially three different risk-based approaches to estimating the ``long-term'' distribution of response (corresponding to hundreds or thousands of years): a) the so-called ``all short-term conditions'' (or ``all sea state'') approach, b) the ``storm event`` approach, and c) the environmental contour method, an approximate method using only short-term analysis. 

There are two distinct joint probability approaches in widespread use in coastal engineering practice in the UK (e.g. \citealt{DEFRA05}). These are (a) a simplified method that involves the use of joint probability contours (JPC) and (b) a risk-based statistical method. Both approaches are implemented within the widely-used JOIN-SEA software system (\citealt{HRW98, HksEA02}). 

\subsection{Environmental contours} \label{Sct:Int:EnvCnt}
%
The environmental contour defines a set of extreme sea state conditions, and can be used to approximate extreme values of long-term structural response extremes by considering only a few short-term metocean conditions. Environmental contours are appealing since they can be specified for a given metocean environment independently of any structure; they are also linked to a well-established approach to structural design, familiar to practitioners. To establish them, joint probabilities of metocean parameters, historically in the metocean community in the form of tables, are needed. The idea behind the method is to define contours in the metocean space (for example, $H_S$, $T_P$) along which extreme responses with given return period should lay. It is a simplified and approximate method compared with full long-term response analysis but requires less computational effort. 

All environmental contour methods have a common goal of summarising the tail of the joint distribution of environmental variables, with a view to learning about the distribution of extreme structural response within a prescribed return period. This is achieved by identifying combinations of environmental variables (sometimes referred to as \textit{governing conditions}) responsible for extreme structural response. Structural responses for combinations of environmental conditions lying on the contour can be used to estimate the extreme response due to a sea state with the same return period as the contour. Importantly, only combinations of metocean parameters lying on the contour need be considered. With additional a priori knowledge of the response, it is possible to limit the interval of the contour over which to evaluate structural response, substantially reducing the computational effort for calculating extreme response. An underlying assumption is that the extreme $N$-year response is governed by sea state conditions on the $N$-year environmental contour.

Contours estimated using different methods will be different in general, since each method makes different assumptions in characterising the environment, or seeks to summarise the environment in a different way. Hence, the choice of environmental contour method will influence the estimation of the distribution of extreme response.

Nevertheless, the environmental contour approach may be useful for early phase concept evaluation. For example, as stated by \cite{NRS17}, if the application under consideration is of a very non-linear nature, an extensive model test program may be necessary to model the short-term variability for all important metocean conditions; the environmental contour approach can help identify those conditions.

Some approaches to estimation of environmental contours (for example IFORM, \citealt{WntEA93}) make additional explicit assumptions regarding the nature of structural failure surfaces expressed in terms of (potentially transformed) environmental variables. When these assumptions are valid, statements regarding the relative magnitude of the exceedance probability of the $N$-year environmental contour and the $N$-year structural failure probability can be made more reasonably. However it is not always clear that the additional assumptions are satisfied for a given application. 

The environmental contour procedure as given by \cite{NRS17} can be summarised as: (a) Establish environmental contours of the metocean characteristics (e.g. $H_S$, $T_P$) corresponding to some annual non-exceedance probability $1-1/T$ (for a $T$-year return period); (b) Identify the worst metocean condition along the contour for the response under consideration; (c) For this sea state, determine the distribution function for the appropriate three-hour (or possible one-hour) extreme value for the response under consideration; and (d) Estimate the value of the response (corresponding to the same annual non-exceedance probability $1-1/T$) using the quantile of the distribution of response (from (c)) with non-exceedance probability $\alpha$. A value $\alpha=0.9$ is recommended for ULS (the ultimate limit state), and $\alpha=0.95$ for ALS (the accidental limit state). The standard provides some guidance as to the adequacy of the approach in terms of the width of the distribution of response in (c). Although this standard discusses the environmental contour method for sea states of length three hours, the procedure can be applied to sea states of any appropriate length (for example, 30 minutes); we use three-hour sea states here for consistency and ease of explanation.

\cite{DNV-RP-C205:2017} recommends two environmental contour approaches: IFORM (\citealt{WntEA93}, procedure similar to \cite{NRS17} above), and a constant probability density approach (\citealt{Hvr87}). The procedure for the latter can be summarised as: (a) Estimate a joint environmental model of sea state variables of interest; (b) Estimate the extreme value for the governing variable for the prescribed return period, and associated values for other variables (for example, 100-year $H_S$ and conditional median $T_P$); and (c) Develop a contour line from the joint model or scatter diagram as the contour of constant probability density going through the parameter combination mentioned above.

Using the environmental contour, an estimate of the extreme response is obtained by searching along the contour for the condition giving maximum characteristic extreme response. The contour method is affected by uncertainties related to metocean data and adopted joint models and has its own limitations which are pointed out by \cite{DNV-RP-C205:2017} and \cite{NRS17}. It will tend to underestimate extreme response levels because it neglects short term variability of response between different realisations of sea states. Both standards recommend approaches based on \cite{WntEng98} to account for this, including (a) increasing the return period corresponding to the contour, and hence \textit{inflating} the environmental contours; (b) replacing the stochastic response by a fixed fractile level higher than the median value; or (c) applying multipliers of the median extreme response estimates, to introduce more conservatism.

In the coastal engineering community, contours of joint exceedance probability of environmental variables are estimated using the JPC method (a) to find design events that form the boundary conditions for numerical and physical models for the purposes of structural design; and (b) to estimate return values of overtopping and overflow rates corresponding to some return period for use in flood mapping and risk analysis (\citealt{CIRIA96, CIRIA07}). A series of combinations of values of environmental variables from the contour are tested in order to find the \textit{worst case} value of the response. This worst case value is then assumed to have the same return period as the return period associated with the environmental contour. Since again, without further assumptions, there is no link between environmental contour and structural response, there are obvious short-comings to this approach, which are well recognised (e.g. \citealt{GldEA17}).

The performance of different environmental contour methods has been investigated in several studies, including work by some of the current authors (including \citealt{JntEwnFln12b, VnmBtn15, GldEA17, Vnm17}). After consideration of the fundamental mathematical differences between different contour methods, it is unreasonable in general to expect to find any consistent trends in comparisons of contour methods across different applications. The characteristics of different environmental contour methods must be assessed on an application-by-application basis. There are a number of more fundamental reviews of environmental contour methods, including excellent recent work by \cite{HslEA17}.

\subsection{End-user survey} \label{Sct:Int:Srv}
%
As part of the ECSADES research project (see Section~\ref{Sct:Ack}), a survey was conducted on end-user practice in the use of environmental contours. The survey, receiving 19 respondents from industry and academia, consisted of questions aimed at establishing the following: (a) Frequency of use of environmental contours in structural reliability and design; (b) Variety in methods used to define the contour and popular sources of guidance and literature; (c) Variety in application of contours (how is information from the contour used); and (d) Perceived advantages and disadvantages of using environmental contours.

Further details of the survey can be found in \ref{SctApp:SrvFnd}. Two key insights resulted from the survey. Firstly, though respondents cited varying frequency of application of contours, they did appear to agree on contours forming an integral part of reliability assessment and design. Respondents cited contours as an ``industry-accepted approach to approximating $N$-year responses quickly'' (when compared to long-term time-domain methods), especially when there is little or no knowledge about the structure being designed - the same environmental contour potentially being applicable to a range of responses.  Secondly, respondents cited that there is no single standard approach to defining the contour nor to applying it in the estimation of responses. Seven key sources of guidelines were cited, part of a collection of over fifty relevant papers collected by the authors of this paper. Further, respondents expressed concern over a lack of understanding of the meaning of the contours and the risks associated with naive application of statistical methods leading to physically unreasonable contours. The level of interest in application offshore is greater in Norway than other locations, although coastal practitioners use contours widely. 

These insights highlight the need for clarity both on the modelling choices available when defining contours, and on the applicability of contours given the information available for a given structure and environment. We have attempted to address the majority of comments emerging from the survey in this paper

\subsection{Objective and outline} \label{Sct:Int:ObjOtl}
%
The objective of this article is to (a) Overview the statistical ideas underpinning environmental contour methods, (b) Highlight fundamental differences between methods, (c) Explain the link between environmental contour and structural failure probability claimed by some approaches, (d) Provide simple software to allow a metocean practitioner to estimate a sensible model for general settings (based on extreme value analysis of a historical sample from the environment), and (e) Provide basic guidance regarding \textit{when} and \textit{how} environmental contour methods should be used sensibly.

The layout of the article is as follows. Section~\ref{Sct:Cnc} discusses fundamental issues regarding the meaning and definition of return value in a multivariate setting, and lack of invariance of a probability density function under transformation of variables. It also discusses the procedure typically used to estimate the distribution of so-called ``long-term'' statistics (such as the $N$-year maximum response) from ``short-term'' statistics (such as the distribution of maximum response in a three-hour sea state). Section~\ref{Sct:MdlEnv} provides a description of different families of models for the joint distribution of environmental variables. Section~\ref{Sct:EstCnt} outlines the different kinds of environmental contours discussed in the literature. It also outlines a rationale to relate the characteristics of an environmental contour with a structural failure surface for some response, and describes an approach to modify environmental contours to account for the short-term variability of maximum response in a three-hour sea state. Section~\ref{Sct:Prc} provides a discussion of case studies used to illustrate the competing characteristics of different environmental contour methods, and the challenges of linking contour with response. Section~\ref{Sct:DscCnc} provides discussion, and a concluding protocol to aid the practitioner in deciding when and how to apply environmental contour approaches reasonably.

\section{Return values, transformation of variables and long-term statistics} \label{Sct:Cnc}
%
In this section, we start by considering the definition of a univariate return value, and consider the issues in extending this concept to the multivariate case. We illustrate the sensitivity of the probability density function for the joint distribution of environmental variables to transformation of variables. We also describe how the characteristics of a response for a sea state (``short-term'') can be used to estimate its characteristics over an extended period of time (``long-term'').

\subsection{Univariate return values} \label{Sct:Cnc:UnvRtrVls}
%
Estimation of extremes for a single variable $X$ is relatively straightforward and has been studied extensively (e.g. \citealt{Cls01}). Given a representative set of independent observations of $X$ spanning many years, extreme value analysis can be used to estimate the distribution $F_{M_{X,1}}$ of the annual maximum $M_{X,1}$. This in turn can be used to estimate the $T$-year return value $x_T$ by solving the equation
\pben
\Pr(M_{X,1}>x_T)=1-F_{M_{X,1}}(x_T)=1/T \quad . \label{E:RV1D}
\peen
The $T$-year return value $x_T$ for a single variable is therefore well-defined in terms of the tail of the cumulative distribution function $F_{M_{X,1}}$ of the annual maximum $M_{X,1}$ of $X$. 

\subsection{Multivariate return values} \label{Sct:Cnc:MltRtrVls}
%
Unfortunately, the joint return value for two or more variables $(X,Y)$ cannot be uniquely defined (e.g. \citealt{Srn15}). For example, in the case of two variables $(X,Y)$, we could define the return value $(x_T,y_T)$ in terms of the joint distribution $F_{M_{X,1},M_{Y,1}}$ of the annual maxima $M_{X,1},M_{Y,1}$ using
\pben
\Pr(M_{X,1}>x_T,M_{Y,1}>y_T)=1-F_{M_{X,1},M_{Y,1}}(x_T,y_T)=1/T \quad . \label{E:RV2D}
\peen
However, it is immediately apparent that there is no unique solution to this equation; given any pair $(x_T,y_T)$ which satisfies the equation, we can increase $x_T$ slightly, and reduce $y_T$ such that the equation is still satisfied. That is, there is a continuum of solutions  to the equation, which we can write as the set $\{x_T(\theta),y_T(\theta)\}$ indexed by parameter $\theta \in \mathcal{C}$. As we vary $\theta$, the pair $(x_T(\theta),y_T(\theta))$ maps out a contour (corresponding to $\{x_T(\theta),y_T(\theta)\}$) of constant exceedance probability $1/T$ in $(x,y)$-space. We note for clarity, that a contour refers to a closed curve in some space ($(x,y)$-space here), on which the value of some function is constant. 

We can rewrite the last equation as
\pbe
\Pr(M_{X,1}>x_T,M_{Y,1}>y_T)=\Pr(M_{Y,1}>y_T|M_{X,1}>x_T)\Pr(M_{X,1}>x_T)=1/T \quad ,
\pee
motivating a definition of return value for the pair in terms of a marginal return value of $M_{X,1}$ and a conditional return value of $M_{Y,1}$ given $M_{X,1}$. Two boundary cases exist: (a) when $M_{X,1}$ and $M_{Y,1}$ are perfectly correlated, points $(x_T,y_T)$ on the solution contour would satisfy $\Pr(M_{X,1}>x_T)=1/T$ since $\Pr(M_{Y,1}>y_T|M_{X,1}>x_T)=1$ in this case, and (b) when $M_{X,1}$ and $M_{Y,1}$ are independent, solutions $(x_T,y_T)$ would satisfy $\Pr(M_{Y,1}>y_T)\Pr(M_{X,1}>x_T)=1/T$ since now $\Pr(M_{Y,1}>y_T|M_{X,1}>x_T)=\Pr(M_{Y,1}>y_T)$. In general, the extent of dependence between pair $M_{X,1}$ and $M_{Y,1}$ will be somewhere between perfect dependence (a) or independence (b).

To illustrate the importance of dependence in practice, we could for example choose to use the 100-year maximum wave height, wind speed and current speed to estimate the environmental loading with a return period of 100 years. If winds, waves and currents are perfectly correlated, the probability of this combination of variables occurring would be $10^{-2}$ per annum as required. But if the variables were independent,  the probability of this combination of variables occurring would be $10^{-6}$ per annum, considerably less than $10^{-2}$. Some design codes and guidelines suggest taking the 100-year return period of one (dominant) variable together with the values of associated variables corresponding to shorter return periods to accommodate dependence between variables (e.g. \citealt{FldEA18}). The DNV recommended practice for on-bottom stability of pipelines suggests the combination of the 100-year return condition for waves combined with the 10-year return condition for current or vice-versa, when detailed information about the joint probability of waves and current is not available.

In summary: the concept of return value is not uniquely defined for more than one variable. In order to specify design values for more than one variable rationally, we need to understand and exploit the joint distribution of the variables. This leads naturally to consideration of joint probabilities, of environmental contours, and of structure variables (such as structural response to environmental loading) which capture the important joint characteristics of (a multivariate) environment in terms of a single ``structure'' or response variable (see Section~\ref{Sct:EstCnt}). 

\subsubsection{Transformation of variables} \label{Sct:Cnc:TrnVrb}
%
Suppose we describe the environment in terms of variables $(X,Y)$ and a distribution with joint density $f_{X,Y}(x,y)$. Suppose further we choose also to describe the environment in terms of transformed variables $(a(X),b(Y))$. Then the corresponding density $f_{a(X),b(Y)}(a(x),b(y))$ is given by
\pbe
f_{a(X),b(Y)}(a(x),b(y)) &=& f_{X,Y}(x,y) \left|\frac{\partial (a(x),b(y))}{\partial (x,y)}\right| \\
&\neq& f_{X,Y}(x,y) \text{ in general\quad,}
\pee 
because of the influence of the final Jacobian term on the right hand side of the first line. As a result, if the set of values $\{x(\theta),y(\theta)\}$ (for $\theta \in \mathcal{C}$, say) yield constant density in $(X,Y)$-space, the transformed set $\{a(x(\theta)),b(y(\theta))\}$ will not do so in $(a(X),b(Y))$-space (if the absolute value of the Jacobian is not unity). Thus, contours of constant probability density on one scale will not be so on a different scale. However, contours defined in terms of cumulative distribution functions are invariant to monotonic transformations of variables: for example, a bivariate contour estimated in terms of $X$ and $Y$ would be equivalent to that estimated in terms of $a(X)$ and $a(Y)$.  

\subsubsection{Distribution of the $N$-year maximum response} \label{Sct:Cnc:Str2Lng}
%
We can formally evaluate the distribution $F_{M_{R,N}}$ of the $N$-year maximum response $M_{R,N}$ (in any 3-hour sea state, e.g. \citealt{JntEwn07a}, \citealt{JntEwnFrr08a}) using
\pbe
F_{M_{R,N}}(r) = \exp\left[-\lambda N \left(1-F_R(r)\right)\right]
\pee 
where $\lambda$ is the expected number of storms per annum, and $F_{R}(r)$ is the distribution of maximum response $R$ in a random storm, given by
\pbe
F_{R}(r) = \int_s \left\{ \int_{\un{x}_1} \int_{\un{x}_2} ... \int_{\un{x}_s} \left[ F_{R|\{\un{X}_i\}}(r|\{\un{x}_i\}) \times f_{\{\un{X}_i\}|S}(\{\un{x}_i\}|s) \right] d\un{x}_1 d\un{x}_2 ... d\un{x}_s \right\} f_S(s) ds
\pee 
where $f_{\{\un{X}_i\}|S}(\{\un{x}_i\}|s)$ is the joint density of sea state variables for a storm of $s$ sea states, and $f_S(s)$ is the density of the number of sea states in a storm. $F_{R|\{\un{X}_i\}}(r|\{\un{x}_i\})$ is the distribution of maximum response in a storm consisting of $s$ sea states with variables $\{\un{X}_i\}_{i=1}^s$, which can be written as
\pbe
F_{R|\{\un{X}_i\}}(r|\{\un{x}_i\}) = \prod_{i=1}^s F_{R|\un{X}_i}(r|\un{x}_i) 
\pee
where $F_{R|\un{X}}(r|\un{x})$ is the distribution of maximum response in a sea states with variables $\un{X}$. If we have access to all of the distributions above, we can use Monte Carlo simulation, numerical integration, importance sampling or similar to estimate $F_{M_{R,N}}$. When this calculation is feasible, we can directly estimate the joint distribution $F_{\un{X}|M_{R,N}}$ of the environmental variables given the occurrence of a $N$-year maximum response (in a 3-hour sea state); these values are often called \textit{associated values} for the environmental variables given the $N$-year response, and can be represented as environmental contours using the methods of Section~\ref{Sct:EstCnt:JntExcCnt}-\ref{Sct:EstCnt:IsoDnsCnt}. However, this calculation can be computationally complex, for example when the evaluation of  $F_{R|\un{X}}$ is demanding, potentially involving time-domain simulation of environment-structure interaction, finite element analysis, etc.; other approximate approaches, including those exploiting environmental contours, are then appealing, as explained in Section~\ref{Sct:Prc}.

In the coastal engineering literature, discussion tends to be in terms of ``risk-based'' estimation as opposed to estimation of ``long term response'', since key concerns are the annual probability of failure or expected annual damage, over a particular epoch (for example,  ``present day'' or ``2050'').

\section{Modelling the joint distribution of environmental variables} \label{Sct:MdlEnv}
%
Given a sample $\{\un{x}_i\}_{i=1}^n$ with $\un{x}_i=\{x_{i1},x_{i2},...,x_{im}\}$ from the joint distribution of $m$ environmental variables $\un{X}$ $=(X_1,X_2,...,X_m)$, a number of different models for the joint distribution have been reported in the literature. Models can be categorised as being parametric (adopting a functional form for the density of the joint distribution) or non-parametric (typically using kernels for density estimation). 

\subsection{Non-parametric models} \label{Sct:MdlEnv:NnP}
%
The simplest form of non-parametric density estimation is kernel density estimation. The joint probability density function $f_{\un{X}}(\un{x})$ of $\un{X}$ evaluated at $\un{x}$ takes the form
\pbe 
	f_{\un{X}}(\un{x})=\sum_{i=1}^n k(\un{x};\un{x}_i,\mathcal{P})
\pee
for $n$ kernel functions $k$ with common parameters $\mathcal{P}$ centred at each of $\{\un{x}_i\}_{i=1}^n$ such that for any $\un{x}'$
\pbe 
	\int_{\un{x}} k(\un{x};\un{x}',\mathcal{P}) d\un{x}=1 \quad .
\pee
A typical kernel choice might be the multivariate normal density $\phi(\un{x};\un{x}',\un{\Sigma})$ with mean $\un{x}'$ and covariance matrix $\un{\Sigma}$. Some of the parameters $\mathcal{P}$ can be set prior to estimation, and the remainder estimated by maximum likelihood estimation. For example, in the case of the multivariate normal density, we might set $\un{\Sigma}=h^2 I_m$ (where $I_m$ is the $m \times m$ identify matrix) so that the model-fitting problem is reduced to estimating a single kernel width parameter $h$. 

Kernel density models are suitable in general for the description of the body of a distribution, and the choice of kernel parameters $\mathcal{P}$ tends not to be too critical to estimate central characteristics. In contrast, kernel density models are not suitable to describe tails of distributions, since the tail (away for locations of observations  $\{\un{x}_i\}_{i=1}^n$) is strongly influenced by the choice of kernel $k$ and kernel parameters $\mathcal{P}$ .

\subsection{Copula models} \label{Sct:MdlEnv:Cpl}
%
Consider a two-dimensional $(H_S,T_P)$ environment. We might try to describe the joint density of $H_S$ and $T_P$ as the product of marginal densities $f_{H_S}(h)$, $f_{T_P}(t)$ for $H_S$ and $T_P$, and a function $\rho_2(F_{H_S}(h),F_{T_P}(t))$ describing their dependence
\pbe
	f_{H_S,T_P}(h,t) = f_{H_S}(h) f_{T_P}(t) \rho_2(F_{H_S}(h),F_{T_P}(t))
\pee 
where $F_{H_S}(h)$ and $F_{T_P}(t)$ are marginal cumulative distribution functions for $H_S$ and $T_P$. $\rho_2$ is the probability density function of a two-dimensional copula, a multivariate probability distribution for which the marginal probability distribution of each variable is uniform. Copula models are useful since they focus on describing the dependence structure between random variables. Before estimating the copula model, we fit marginal distributions to $H_S$ and $T_P$; tails of marginal distributions can be estimated using extreme value models. In general for set $\un{X}=\{X_1,X_2,...,X_m\}$ of environmental variables, the copula density takes the form
\pbe
	f_{\un{X}}(\un{x}) = \left[\prod_{j=1}^m f_{X_j}(x_j)\right] \rho_m(F_{X_1}(x_1), F_{X_1}(x_2), ..., F_{X_m}(x_m))
\pee 
where $\rho_m$ is the density of an $m$-dimensional copula. There is a huge literature on copulas (e.g. \cite{Nls06}, \citealt{JwrEA10}), and there are many families of copulas (including the Gaussian and Archimedian), and some (so-called \textit{max-stable} or inverted max-stable copulas) more suited to the descriptions of extreme environments.  \cite{DrtSlv09} illustrate the specification and estimation of multivariate extreme value models using copulas. \cite{GdnSgr10} discusses the special class of extreme value (or max-stable) copulas appropriate for describing joint tails of distributions of component-wise maxima. \cite{RbtMhm12} provides an excellent review of extreme value copulas and their relationship to max-stable processes. Copula methods have received some attention in the ocean engineering literature. \cite{FzrFrrEA18} discuss the use of copulas in metocean design.  \cite{MntItrHrdZvn15, MntItrHrdZvn16, MntItrHrdZvn17} discuss the estimation of environmental contours using copula methods. \cite{BndEA14} and \cite{SrfRgg14} propose bivariate extreme value models incorporating non-stationary marginal and dependence inference. Asymmetric copula models were found to be necessary to model $H_S$, $T_Z$ by \cite{Vnm16a}.

\subsection{Hierarchical conditional models} \label{Sct:MdlEnv:Hrr}
%
In a hierarchical model, the structure of the dependence between environmental parameters takes on a particularly advantageous form. Again consider the case of $H_S$ and $T_P$; the joint density $f_{H_S,T_P}(h,t)$ can be written in the form
\pbe
	f_{H_S,T_P}(h,t)=f_{T_P|H_S}(t|h)f_{H_S}(h) .
\pee
It is always possible to factorise the joint density into the product of densities. For $H_S$ and $T_P$,  because of their physical characteristics, the densities $f_{T_P|H_S}$ and $f_{H_S}$ are relatively simple: a Weibull distribution for $H_S$ has been used for many years, and a log-normal distribution for $T_P|H_S$ (e.g. \cite{BtnHvr89}, \cite{MthBtn90}, \citealt{HvrWnt09}). These are combined to estimate the joint density $f_{H_S,T_P}$. More generally, for example in the case of three environmental variables $\{X_1,X_2,X_3\}$, it is always possible to factorise the joint density as
\pbe
f_{X_1,X_2,X_3}(x_1,x_2,x_3)=f_{X_3|X_1,X_2}(x_3|x_1,x_2)f_{X_2|X_1}(x_2|x_1)f_{X_1}(x_1)
\pee
with equivalent factorisations for permutations of the three variables. Estimating the joint distribution on the left hand side therefore reduces to estimating all of the distributions on the right hand side. Depending on the statistical characteristics of $\{X_1,X_2,X_3\}$, estimating all the distribution on the right hand side may be more straightforward to achieve in practice, in which case the factorisation is useful. Specifying a physically-realistic and useful conditional structure for $m$ variables $\{X_1,X_2,...,X_m\}$ becomes increasingly problematic as $m$ increases; \cite{BtnHvr89, BtnHvr91}, \cite{FrnEA07}, \cite{Btn15} provide examples with different levels of complexity of dependence structure. The conditional structure can often be usefully expressed as a graphical model (e.g. \citealt{Jrd04, Brb12}). 

Once a useful conditional structure is established, we need to estimate the densities involved. In general, different functional forms are considered based on inspection of the data. For the tail of the distribution of a random variable, a standard tail distribution (for example, Weibull, Gumbel, Frechet, generalised Pareto, generalised extreme value) would seem to be a reasonable choice. Choice of a suitable generic density for conditional densities for (say) $X_2|X_1$ or $X_3|X_2,X_1$ is less obvious. \cite{BtnHvr89, BtnHvr91} propose a two-parameter Weibull distribution to describe the conditional distribution of wind speed given $H_S$. \cite{HrnEA18} applies the approach to specify the joint distribution of a relatively large number of environmental variables. Ideas from hierarchical graphical and copula models can be combined, as illustrated by \cite{YuDwlJnt14} in a metocean context.

\subsection{Conditional extremes model} \label{Sct:MdlEnv:Cnd}
%
The conditional extremes model is motivated by the existence of an asymptotic form for the limiting conditional distribution of one or more conditioned random variables given a large value of a conditioning variable for a large class of distributions (e.g. \citealt{HffTwn04, HffRsn07}), on particular standard marginal scales. For a set $\un{X}=\{X_1,X_2,...,X_m\}$ of environmental variables, it provides a flexible framework to estimate the joint distribution of $\un{X}_{-k}=\{X_1,X_2,...,X_{k-1},X_{k+1},...X_m\}$ given that $X_k$ ($k=1,2,...,m)$ is extreme in its marginal distribution. The modelling procedure proceeds in four steps: (a) marginal extreme value modelling of each of $\{X_j\}$ independently, followed by (b) marginal transformation of each of $\{X_j\}$ independently to the corresponding variable in $\{\tilde{X}_j\}$ with standard Laplace marginal distribution, (c) dependence modelling of $\un{\tilde{X}}_{-k}|\tilde{X}_k>\psi$ for large $\psi$ for each $k$, and (d) simulation under the estimated model to estimate return values, environmental contours, etc. The conditional extremes model with parameters $\un{\alpha}_{-k} \in [0,1]^{m-1}$ and $\un{\beta}_{-k} \in (-\infty,1]^{m-1}$ is given by
\pbe
\un{\tilde{X}}_{-k}|\{\tilde{X}_k=\tilde{x}_k\} = \un{\alpha}_{-k} \tilde{x}_k + \tilde{x}_k^{\un{\beta}_{-k}} \un{W}_{-k}  \text{ for } \tilde{x}_k>\psi, \text{ for each } k
\pee
where $\un{W}_{-k}$ is a residual process assumed to be distributed as $\un{W}_{-k} \sim \texttt{MVN}(\un{\mu}_{-k}, \texttt{diag}(\un{\zeta}_{-k}))$ with mean $\un{\mu}_{-k}$ and variance $\un{\zeta}_{-k}$ (elements of which positive) for model estimation only. Threshold $\psi \in \mathbb{R}$ is defined as the quantile of the standard Laplace distribution with appropriately high non-exceedance probability $\kappa \in (0,1)$. \cite{KefPpsTan13} provide additional constraints on the parameters of the conditional extremes model. An outline of the approach is given by \cite{JntFlnEwn10} in application to wave spectral characteristics.

Extensions incorporating covariates (\citealt{JntEwnRnd14}), to conditioning on multiple locations (\citealt{PpsEA17}), to modelling the evolution of time-series (\citealt{WntTwn16, TndEA18}) and the spatial distribution of extremes (\citealt{ShtEA18}) have recently been reported. The main advantage of the conditional extremes model compared with copula or hierarchical models is that it incorporates a full class of asymptotic extremal dependence (e.g. \cite{ClsHffTwn99}), and also allows relatively straightforward extension to higher dimensions. Of course, the conditional extremes method relies on having sufficient sample to be able to estimate the marginal and conditional tails adequately.

\section{Estimating contours} \label{Sct:EstCnt}

In this section, we outline the different types of contour (in Sections~\ref{Sct:EstCnt:JntExcCnt} and \ref{Sct:EstCnt:IsoDnsCnt}) typically used to describe the environment, and then discuss how  response-based design (Section~\ref{Sct:EstCnt:Cnt2Rsp}) also yields joint distributions of \textit{associated} environmental variables which can be usefully summarised using a contour. First, we provide a brief summary of some of the literature on joint modelling of the ocean environment leading to contour estimation.

\subsection{Overview of literature} \label{Sct:EstCnt:Int}
%
Joint modelling of environmental variables, and the construction of environmental contours, has a long history. \cite{Hvr87}, \cite{BtnHvr89, BtnHvr91} present joint models for environmental variables from which environmental contours can be estimated. \cite{WntEA93} introduces the IFORM method, motivated by transformation of the joint distribution of environmental variables to standard multivariate Normal using the Rosenblatt transformation.

\cite{Ler08} provides a comparison of stochastic process models for definition of design contours. \cite{JntEwnFln12b} and \cite{HsbEA15a} present methods for estimation of joint exceedance contours based on direct Monte Carlo simulation under a model for the joint distribution of environmental variables. \cite{HslEA17} estimate highest density contours, again using a random sample simulated under a model for the joint distribution of environmental variables. \cite{HslEA17} also provides an illuminating discussion of the characteristics of different approaches to environmental contour estimation. The approaches of \cite{Hvr87}, \cite{JntEwnFln12b} and  \cite{HslEA17} seek only to find contours which describe the distribution of environmental variables. The methods of \cite{WntEA93} and \cite{HsbEA15a}, with extra assumptions, provide a direct link between the characteristics of the environmental contour and structural failure. \cite{VnmBtn15} and \cite{Vnm17} provide comparisons of different approaches to contour estimation. 

Other literature (e.g. \citealt{TrmVnd95}, \citealt{GldEA17}, \citealt{JnsEA18}) discusses how joint models for the environment can be combined with simple models for structural responses given environment, to estimate the characteristics of response directly. As a result, the joint distribution of environmental variables corresponding to an extreme response can be estimated. \cite{Wnt15a} discusses incorporating the effects of direction and other sources of non-stationarity or homogeneity in design contour estimation. More generally, it is interesting also to consider the influence of different structural responses (and modes of failure) active for a particular structure on the desired characteristics of corresponding environmental contours. For example, if there are $n_R$ \textit{independent} structural responses, and the structure designed so that the probability of failure with respect to each response is $p_F$, then the overall failure probability is $n_R p_F$; yet if the responses are perfectly correlated the overall failure probability is still only $p_F$. It would seem reasonable in general to design to the overall failure probability, and to adjust failure probabilities for individual responses to account for dependence; this in general would also require inflation of environmental contours. We note recent developments which seek to estimate buffered environmental contours (\citealt{DhlHsb18}) which incorporate not just structural failure, but the extent of structural failure.

Because of its prevalence in ocean engineering practice, we start our overview of methods for contour estimation using the IFORM approach.  Then (in Section~\ref{Sct:EstCnt:JntExcCnt} and Section~\ref{Sct:EstCnt:IsoDnsCnt}) we describe related approaches to estimating joint exceedance and isodensity contours. Finally, we consider direct estimation of the distribution of long-term response (in Section~\ref{Sct:EstCnt:Cnt2Rsp}).

\subsection{IFORM contours} \label{Sct:EstCnt:IFORM}
%
The IFORM method of \cite{WntEA93} typically assumes a hierarchical model (Section~\ref{Sct:MdlEnv:Hrr}) for the joint distribution of the environmental variables. We assume that we can describe the joint distribution of variables $\Xb$ sufficiently well that a transformation of variables is possible, so that the joint probability distribution of the transformed variables takes on a particularly simple form. The transformation is achieved by re-expressing the set $\Xb =\{X_1, X_2, ..., X_m\}$ as a set of (independent) conditional random variables $\Xtb=\{\Xt_1, \Xt_2, ..., \Xt_m\}$. For example, for appropriately ordered variables we can write $\Xt_1=X_1$, $\Xt_2=X_2 | X_1$ and $\Xt_3=X_3 | X_1, X_2$  and so on, with cumulative distribution functions $F_{{\Xt}_1}, F_{{\Xt}_2}, \dots$ such that
\pbe
F_{\Xb}(\xb)=\prod_{j=1}^m F_{\Xt_j}(x_j) \quad .
\pee
That is, by design, the random variables $\Xtb$ are independent of each other, and can hence be transformed independently to standard Gaussian random variables $\Ub$ $=\{U_1, U_2, ...,U_m\}$ via the probability integral transform
\pbe
F_{\Xt_j}(x) = \Phi(u_j) \text{ for } j=1,2,...,m
\pee
where $\Phi$ is the cumulative distribution function of the standard Gaussian distribution. Isodensity contours in $\un{U}$-space, with a given non-exceedance probability, can be back-transformed to the original physical space. It should be noted (Section~\ref{Sct:Cnc:TrnVrb}) that isodensity in $\un{U}$-space does not correspond to isodensity in $\un{X}$-space however. The non-exceedance probability corresponding to the IFORM contour can be related to the probability of structural failure, given certain assumptions, as explained in Section~\ref{Sct:EstCnt:Cnt2Rsp}. We also note recent work by \cite{WeiBrn18} on constructing inverse second-order (ISORM) contours.

\subsection{Joint exceedance contours} \label{Sct:EstCnt:JntExcCnt}
%
The equations in Section~\ref{Sct:Cnc} illustrate the key characteristic of a $T$-year return value: namely that it defines a region $\mathcal{A}$ with closed boundary $\{\un{x}(\theta)\}$ for $\theta \in \mathcal{C}$ of the domain over which variables are defined associated with probability $1-1/T$ per annum, and a complementary set with ``exceedance'' probability $1/T$. In a multivariate setting, for a pair of variables for simplicity, Equation~\ref{E:RV2D} shows one way to define $\mathcal{A}$ using the joint cumulative distribution function, leading to the so-called \emph{joint exceedance} contour. However, $\mathcal{A}$ could be defined quite arbitrarily, provided that it corresponds to the desired non-exceedance probability. In practice, $\mathcal{A}$ might even correspond to the union of disjoint sets; the only requirement is that the probability $p$ associated with $\mathcal{A}$ is $1-1/T$. 

The IFORM procedure of Section~\ref{Sct:EstCnt:IFORM} provides a specific approach to the estimation of region $\mathcal{A}$ and hence of joint exceedance contours. IFORM is used typically for offshore applications. Joint exceedance contours are also widely used in coastal applications(e.g. \citealt{DEFRA03}, \citealt{GldEA17}), and their limitations in terms of naive estimation of extreme responses recognised for some time.

\subsubsection{Direct sampling contours (\citealt{HsbEA15a})} \label{Sct:EstCnt:JntExcCnt:DrcSmpCnt}
%
The IFORM-method (Section~\ref{Sct:EstCnt:IFORM}) produces a contour where the probability of any convex failure region in the transformed Gaussian $\un{U}$-space which do not overlap with the interior of the contour is less than or equal to a given target probability. When this contour is transformed back to the environmental $\un{X}$-space, however, this probabilistic interpretation is no longer valid (as explained in Section~\ref{Sct:Cnc:TrnVrb}). The direct sampling contour (\citealt{HsbEA13, HsbEA15a, HsbEA15b, HsbEA17}) is constructed so that it has the same probabilistic properties in the environmental space as the IFORM contour has in transformed space. This implies that the region $\mathcal{A}$ enclosed by the direct sampling contour is always convex; that is, for any two points in $\mathcal{A}$, the straight line joining them would also be in $\mathcal{A}$. 

Estimating the direct sampling contour in two-dimensions is relatively easy, based on a simulation under a model for the joint distribution of variables $X_1$ and $X_2$ (see Section~\ref{Sct:MdlEnv}). For probability level $\alpha$, following \cite{HsbEA15a}, we first find the function $C(\theta)$, the $(1-\alpha)$-quantile of the distribution of the projection $X_1 \cos(\theta)+X_2 \sin(\theta)$ for each value of $\theta\in[0,2\pi)$
\pbe
	C(\theta)=\inf\left\{C:\Pr\left(\left[X_1 \cos(\theta)+X_2 \sin(\theta)\right])>C\right)=\alpha\right\} .
\pee
Then we estimate the contour $\mathcal{C}=\{(x_1(\theta),x_2(\theta)) : \theta \in [0,2\pi)\}$ using
\pbe
	x_1(\theta) &=& C(\theta) \cos(\theta) - \frac{dC}{d\theta}\sin(\theta) , \\
	x_2(\theta) &=& C(\theta) \sin(\theta) + \frac{dC}{d\theta}\cos(\theta) ,
\pee
and potentially further smooth $\mathcal{C}$ as a function of $\theta$. Following \cite{WntEA93}, for at $T$-year return period, it is recommended that the value of $\alpha$ be set to $1/T$. Generalisations to higher dimensions is mathematically straightforward; three-dimensional contours based on the direct sampling approach are presented in \cite{Vnm18}. 

\subsubsection{Joint exceedance contours (\citealt{JntEwnFln12b})} \label{Sct:EstCnt:JntExcCnt:JntEA}
%
\cite{JntEwnFln12b} propose joint exceedance contours for which a particular probability at any point on the contour is constant. Specifically, in two-dimensions, the closed contour $\{(x_1(\theta),x_2(\theta)) : \theta \in [0,2\pi)\}$ is defined by 
\pbe
 \Pr(\bigcap_{j=1}^2 (r_j(\theta; r^*) X_j > r_j(\theta; r^*) x_j(\theta))) = \alpha
\pee
for $\alpha \in (0,1)$, where $r(\theta; r^*) = \{r_1(\theta; r^*), r_2(\theta; r^*)\}$ is defined by $r(\theta; r^*) = x(\theta) - r^*$ and $r^*$ is a reference location for the distribution under consideration. In some cases, it is appropriate that $r^*$ refer to some central feature (for example, mean, median or mode). In other situations, when we are interested solely in the large values of a variable $X_1$ (say), it might be appropriate to set $r_1^* = 0$. We estimate the contour using simulation under a model for the joint distribution of $X_1, X_2$; but we might choose to estimate the contour for any transformation of variables, in particular to independent standard normals $U_1,U_2$. The probability $p$ associated with region $\mathcal{A}$ enclosed by the contour is a (generally unknown) function of $\alpha$. The value of $p$ can be set to approximately $1-1/T$ by iteration over different choices of $\alpha$. Again,  we can potentially further smooth the contour as a function of $\theta$.

\subsection{Isodensity contours} \label{Sct:EstCnt:IsoDnsCnt}
%
Another obvious approach would be to define contours using the joint probability density function $f_{\un{X}}$ of the environmental variables instead of its cumulative distribution function $F_{\un{X}}$. If we assume to start that $f_{\un{X}}$ is uni-modal, we might choose a value $\tau$ such that
\pbe
f_{\un{X}}(\un{x}_\tau)=\tau
\pee
defines a closed contour $\{\un{x}_\tau(\theta)\}$ for $\theta \in \mathcal{C}$ enclosing a set $\mathcal{A}$ in $\un{x}$-space such that $f_{\un{X}}>\tau$ within $\mathcal{A}$. This defines an \emph{isodensity} contour (or a contour of constant probability density); see e.g. \cite{DNV-RP-C205:2017}. The approach can be extended to include multi-modal $f_{\un{X}}$ (\citealt{Hvr87}, \citealt{HslEA17}).

Kernel density estimation (Section~\ref{Sct:MdlEnv:NnP}) is a popular choice for estimation of isodensity contours, but this choice is problematic for estimating the tails of distributions since the relatively arbitrary choice of kernel function and its width tend to dominate the shape of the tail; we would prefer that the shape of the tail was informed more directly by the data. 

\subsection{Relating environmental contours to response} \label{Sct:EstCnt:Cnt2Rsp}
%
If the objective of a study is to estimate environmental conditions corresponding to extreme structural responses, the obvious approach is direct estimation of the characteristics of the $N$-year maximum response and the environments which generate it. In contrast to Section~\ref{Sct:EstCnt:JntExcCnt}-\ref{Sct:EstCnt:IsoDnsCnt}, the purpose of this analysis is not characterisation of extreme environments, but rather of environments related to extreme responses. Response-based methods obviously require at least some information about the response function. Of course, if structural response is monotonically related to a dominant environmental driver variable, then the resulting contours may be quite similar; but this is not always the case. A number of different approaches relying on some knowledge of the response have been developed and applied over the past thirty years (e.g. \cite{ClsTwn90, Twn92, WntEA93, TrmVnd95, HksEA02}).  

\subsubsection{Approximating the response} \label{Sct:EstCnt:Cnt2Rsp:App}

When evaluation of $R|\un{X}$ is demanding (see Section~\ref{Sct:Cnc:Str2Lng}), an alternative approach is to adopt a simple approximation to the response function which can be easily evaluated. Sometimes, the functional form is relatively apparent from physical considerations (for example, the semi-empirical Morison equation (\citealt{MrsEA50}) for the drag and inertial forces on a body); in this case, it is usually necessary to set the parameters of the response function to correspond with the structure of interest. Nevertheless, once set, structural loads can be estimated quickly for given environment $\un{x}$. This is the basis of, for example, the \textit{generic load model} of \cite{TrmVnd95}, and the response-based joint probability model in coastal applications (e.g. \citealt{GldEA17}). More generally, a statistical model (known as an \textit{emulator}) can be used to estimate $R|\un{X}$. The emulator is estimated by evaluating $R|\{\un{X}=\un{x}\}$ for points $\un{x}$ drawn from a set of representative environments $\un{X}$ (which itself can be a computationally demanding analysis), and then fitting a statistical model such as a response surface to explain response in terms of the environmental variables. Once estimated, the emulator provides rapid evaluation of $R|\{\un{X}=\un{x}\}$ for any $\un{x}$, and hence of associated environmental values corresponding to the $N$-year maximum response (in Section~\ref{Sct:Cnc:Str2Lng}). It is a natural framework for error propagation and uncertainty quantification.

\subsubsection{Sampling from environmental contours} \label{Sct:EstCnt:Cnt2Rsp:DrcSmpIFORM}
%
The direct estimation of the distribution $F_{M_{R,N}}$ of the $N$-year maximum response $M_{R,N}$, and hence of the joint distribution $F_{\un{X}|M_{R,N}}$ of environmental variables associated with it is often computationally complex. In such cases, reducing the number of evaluations of $R|\un{X}$ made is advantageous. It is intuitive therefore that we should focus on values $\un{x}$ of environmental variables (to evaluate $R|\{\un{X}=\un{x}\}$) corresponding to extreme environments to achieve this; the $N$-year environmental contour provides one approach to identifying those environments. We can also associate the $N$-year environmental contour with the probability of structural failure in the same period, as described in Section~\ref{Sct:Int:EnvCnt} and outlined here.

\subsubsection*{Reliability theory}
%
Engineering codes stipulate that marine structures should be designed to exceed specific levels of reliability, usually expressed in terms of an annual probability of failure. Reliability theory provides an approach to estimating required structural strength and environmental design conditions causing failure. Structural failure is assumed to occur when structural loads $R$ exceed structural resistance or strength $S$, expressed in terms of the equation
\pbe
S-R=g_{\Xb}(\Xb)<0 \quad ,
\pee
where $g_{\Xb}(\Xb)$ is a limit state expression for a particular failure mechanism (or ``failure surface'' for brevity) and the vector $\Xb$ represents all of environmental, hydraulic loading and structural variables appropriate to a particular problem, with joint probability density function $f_{\Xb}(\xb)$. In the context of floating structures and ships, $R$ might correspond to a motion response of the vessel (such as roll) and $S$ to a limiting value for roll at which structural integrity is deemed impaired. The probability $p_F$ of structural failure can then be evaluated using
\pben
p_F = \Pr(g_{\Xb}(\Xb) < 0) = \int_{g_{\Xb}(\xb) < 0} f_{\Xb}(\xb) \ d\xb \quad .
\label{E:pF}
\peen
For a given environment, $p_F$ can clearly be reduced by increasing $S$, since then the region of $\xb$ space for which $S-R<0$ is reduced. In this way, structural strength can be adjusted to achieve desired $p_F$. Solving Equation~\ref{E:pF} however presents multiple challenges of first specifying $g_{\Xb}$ and $f_{\Xb}$ adequately, and then performing the computation reasonably. We note that the form of $g_{\Xb}$ is in general quite arbitrary, and that estimating $g$ adequately for a full-scale structure is likely to be problematic. If we suspect that extreme environments $\Xb$ produce extreme responses $R$, adequate characterisation of the joint tails of $f_{\Xb}$ will be necessary to estimate $p_F$ well, requiring careful multivariate extreme value analysis of the environment. However, for a resonant frequency response of a floating structure, estimating some tail aspects may be less critical. Other responses such as fatigue are not governed by extremes of the environment in general.

\subsubsection*{Linearising the failure surface}
%
Due to the complexity of solving Equation~\ref{E:pF}, approximate approaches have been sought, including the use of environmental contours. For example, IFORM typically adopts a hierarchical model outlined in Section~\ref{Sct:EstCnt:IFORM} to estimate $f_{\Xb}$ and hence contours of constant probability density in the transformed Gaussian $\un{U}$-space, with given probabilities of non-exceedance. The direct sampling method of Section~\ref{Sct:EstCnt:JntExcCnt:DrcSmpCnt} generates contours in the original $\un{X}$-space of environmental variables with given non-exceedance probability.

Solving Equation~\ref{E:pF} is still not easy, since failure surface $g_{\un{X}}$ is unknown. To overcome this, the direct sampling and IFORM methods make the assumption that a linear approximation to the failure surface at the design point is appropriate. This approximation is made on the original $\un{X}$-scale for direct sampling, and on the transformed $\un{U}$-scale for IFORM; this is the key difference between the methods. There is no a-priori physical reason for assuming that linearisation of the failure surface is appropriate, and the assumption must be justified on engineering grounds on a case-by-case basis. In certain applications, for example of wave loads on fixed structures, extreme loads typically correspond to severe sea states; in this situation, we might assume load to be dominated by significant wave height $H_S$. It is then probably reasonable to assume that the set $\un{x}$ of values (including $H_S$) such that $g_{\Xb}(\un{x})<0$ (and the corresponding set $\un{u}$ of values such that $g_U(\un{u})<0$) is \textit{convex}. In this case, assuming $g_{\Xb}$ (or $g_{\Ub}$) to be linear leads to a \textit{conservative overestimate} of the probability of failure associated with a given contour. However, we emphasise that there is no guarantee that either $g_{\Xb}$ or $g_{\Ub}$ is convex in general. Hence there is no guarantee that linearising the failure surface is reasonable, and that the probability of failure will be smaller than that associated with the environmental contour.

\subsubsection*{Finding governing conditions}
%
In a typical IFORM analysis, once the environmental contour $\{\un{u}(\theta)\}$ (the surface of a hypersphere, for $\theta \in \mathcal{C}$) is estimated in $\un{U}$-space, we find the point $\un{u}(\theta^*) \in \{\un{u}(\theta)\}$ corresponding to the largest values of response (and hence the lowest value of structural failure $p_F$ for given structural strength $S$). Then $\un{u}(\theta^*)$ is transformed back to a corresponding $\un{x}(\theta^*)$ (in terms of the original variables) which is taken as the design set corresponding to the specified failure probability. Other points of interest, for example the whole contour $\{\un{u}(\theta)\}$ in the transformed space, can be similarly transformed to the original space. In a direct sampling analysis, the point $\un{x}(\theta^*)$ can be identified directly in $\un{X}$-space.

\cite{LtsWnt14} generalised IFORM to dynamic systems. \cite{ChaLer18} describe the use of the second-order reliability method (SORM) for contour estimation. 

\subsection{Adjusting contours for model mis-specification and short-term variation} \label{Sct:EstCnt:Adj}
%
We use the $N$-year environmental contour for the set $\un{X}$ to provide a computationally-fast but potentially biased estimate of the $N$-year response for the structure discussed in Section \ref{Sct:EstCnt:Cnt2Rsp}. A typical approach is to estimate the distribution of the maximum response (in any 3-hour sea state, corresponding to a specified return period) given values of environmental variables $\{\un{x}(\theta)\}$ on the contour to identify a ``design point'' $\un{x}(\theta^*) \in \{\un{x}(\theta)\}$ yielding the largest structural response. Then a quantile $q_C$ with non-exceedance probability $p_C$ (for example typically the mode, median or mean) of the distribution $F_{R|\un{X}}(r|\un{x}(\theta^*))$ is used to estimate some (possibly different) quantile $q_R$ with non-exceedance probability $p_R$ of the distribution $F_{M_{R,N}}(r)$ of the $N$-year maximum response. Quantile $q_C$ is used as an estimate for $q_R$, where
\pbe
F_{R|\un{X}}(q_C|\un{x}(\theta^*)) = p_C
\pee
and
\pbe
F_{M_{R,N}}(q_R) = p_R \quad .
\pee
There is no guarantee that $q_C$ and $q_R$ will coincide. Even if $p_C=p_R$, since $F_{R|\un{X}}(r|\un{x})$ has a long right-hand tail for any environment $\un{x}$, it is usually possible for ``short-term variation'' in ``less extreme'' sea states to contribute to the distribution $F_{M_{R,N}}$, whereas this is by definition not possible for the corresponding distribution $F_{R|\un{X}}(q_C|\un{x}(\theta^*))$ from a single sea state $\un{x}(\theta^*)$. Mis-specification of the environmental model, or violation of assumptions concerning the relationship between environment and response, may also lead to disagreement between $q_C$ and $q_R$. For this reason, it is useful to define an inflation factor $\Delta$ such that 
\pbe
q_R = \Delta q_C,
\pee
where we might expect $\Delta>1$ for $p_C \approxeq p_R$. The factor $\Delta$ can be used to inflate the whole environmental contour if desired. Standard \cite{DNV-RP-C205} makes recommendation for appropriate choices of $q_R$, $q_C$ and the corresponding $\Delta$. Section~\ref{Sct:Prc} provides illustrations of contour adjustment for simple simulation models. It is apparent that, in situations where short-term variability is relatively large compared with long-term variability, contour-based approaches should be use with great caution.

\section{Case studies: contours in practice} \label{Sct:Prc}
%
Section~\ref{Sct:EstCnt} outlines various forms of an environmental contour. In the absence of a unified approach to defining and applying contours (see Section~\ref{Sct:Cnc} and Section~\ref{Sct:Int:Srv}), it is informative to consider the practicalities of environmental contour estimation. Our objective in Section~\ref{Sct:Prc} is to quantify how well estimates of extreme responses (in a three-hour sea state, for a particular return period) on a contour compare with estimates obtained by direct simulation of the response. In this sense we replicate the typical approach to application of contours: looking at response for a small set of environmental conditions, in the hope that this analysis approximates the characteristics of maximum response for that return period. In doing so, we discuss the key challenges in applying contours, including choice of contour, sampling along the contour and contour inflation.  As opposed to typical applications, we perform our analysis for four responses, whose relationship to the environment is quantified entirely in terms of $H_S$ and $T_P$, and is known to us. We are therefore able to simulate from the known distributions to estimate the correct characteristics of response, and hence to quantify the performance of contour-based estimated for maximum response. Then, in Section~\ref{Sct:DscCnc}, we summarise our findings regarding the estimation and application of environmental contours for metocean design, with a particular focus on the appropriate use of contours, given the extent of knowledge about the response: environmental contours are clearly useful under certain conditions, but these conditions need to be carefully defined so that the user knows when environmental contours are likely to be a good option.

For simplicity in the case studies below, we define the environment in terms of a large historical sample of sea-state $H_S$ and $T_P$ for a typical northern North Sea environment for the period 1979-2013, from the NORA10-WAM hindcast (\citealt{RstEA11}). NORA10 (Norwegian ReAnalysis 10km grid) is a 58-year hindcast that has been developed by the Norwegian Meteorological Institute. It is a regional HIRLAM (atmosphere) and WAM Cycle-4 (wave) hindcast covering Northern European waters. The regional model uses wind and wave boundary conditions from the ERA-40 reanalysis (1958-2002) and is extended using the ERA-Interim reanalysis from 2002 onwards. NORA10 produces three-hourly wave and wind fields at 10km resolution. We isolate storm peak events using the procedure of \cite{EwnJnt08}. We then estimate structural responses using \textit{known} non-linear functions of environmental variables corresponding to each storm event. 

To construct an environmental contour, we require a statistical model for the environment. Here, we achieve this by means of a conditional extremes model (Section~\ref{Sct:MdlEnv:Cnd}) for the historical sample, using a penalised piecewise constant (\texttt{PPC}) extreme value model (\citealt{RssEA17b}) and software (outlined in \ref{SctApp:PPC}). We choose the conditional extremes model because of its generality and flexibility to model different forms of extremal dependence (e.g. \citealt{JntFlnEwn10}). The \texttt{PPC} extreme value model allows the estimation of non-stationary marginal and conditional extremes for peaks over threshold using a simple description of non-stationarity with respect to covariates in marginal and dependence models. We use the \texttt{PPC} model to estimate a number of the environmental contours discussed in Section~\ref{Sct:EstCnt} and investigate their characteristics, in particular their relationship to extremes of structural response. Because of its recent popularity, we consider the direct sampling contour (Section~\ref{Sct:EstCnt:JntExcCnt:DrcSmpCnt}) in case studies 1 and 2. In case study 2, we also consider the joint exceedance contour outlined in Section~\ref{Sct:EstCnt:JntExcCnt:JntEA} and the isodensity contour (Section~\ref{Sct:EstCnt:IsoDnsCnt}).  To estimate any of these contour methods requires a ($H_S$, $T_P$) sample simulated under the environmental model. The isodensity contour is similar to the approach of \cite{Hvr87} recommended in the \cite{DNV-RP-C205:2017} standard. 

\subsection{Case study 1} \label{Sct:Prc:Cas1}
%
In this case study we consider the direct sampling contour of Section~\ref{Sct:EstCnt:JntExcCnt:DrcSmpCnt} only. The objective of the case study is to examine the general correspondence between estimates for the distribution of the $100$-year maximum response $M_{R,100}$. We compare an estimate from direct simulation of $R$ (taken to be accurate) and one generated from combinations of $H_S$ and $T_P$ on the $100$-year environmental contour. We make this comparison for a number of different responses. 

The procedure we use is intended to reflect common practice in industry. Once the contour is estimated, we identify a ``frontier'' interval of the contour which we think might be informative for estimation of response. In the current work, we assume that the ``frontier'' corresponds to the whole interval of the environmental contour lying close to pairs of $H_S$, $T_P$ values present in the sample. Then we consider two possibilities: (a) that only a single combination of $H_S$ and $T_P$ corresponding to the maximum value $H_S^{\text{max}}$ of $H_S$ on the contour is informative for estimating $M_{R,100}$, and (b) that the whole frontier interval is informative. Then, for scenario (a), we estimate the distribution of maximum $100$-year response $f_{M_{R,100}\text{Point}}(r) = f_{R|\un{X}}(r|H_S^{\text{max}})$. For scenario (b), we estimate the ensemble distribution
\pbe
	f_{M_{R,100}\text{Frontier}}(r) = \frac{1}{n_L}\sum_k f_{R|\un{X}}(r|\un{x}_{100}(\theta_k)) 
\pee
where $\{\un{x}_{100}(\theta_i)\}_{k=1}^L$ defines a set of equally-spaced points on the 100-year contour on the frontier interval. We can then compare quantiles of the distributions from scenarios (a) and (b) with quantiles estimated from direct simulation of $M_{R,100}$. \cite{Vnm17} notes a trade-off between the number of points on the contour used to evaluate the response, the quality of the estimate of response and the computation time required.

A total of four responses $R_1$, $R_2$, $R_3$, $R_4$ were considered. Two responses correspond to output of a structural response simulator for maximum base shear ($R_1$, for a typical fixed structure) and maximum heave ($R_2$, for a floating structure), as a function of $H_S$ and $T_P$ for a three-hour sea state. These response simulators assume that the most probable value of maximum response in a sea state can be written as a closed form expression in terms of a number of sea state variables, including sea state $H_S$ and $T_P$. The actual value of maximum response is then simulated from a Rayleigh distribution with the most probable maximum response as scale parameter. 

A further two synthetic responses are defined, which are simple deterministic functions of $H_S$ and $T_P$, using the following equation
\begin{align}
	R_{i} = \frac{\alpha_{i}H_{S}}{(1+\beta_{i}(T_{p}-T_{p0,i})^2)} \text{ for } i=3,4,
\end{align}  \label{Eqn:SynRsp}
where $T_{p0,i}$ (in seconds) is the resonant peak period for response $R_i$ . The values of $\{\alpha_i,\beta_i,T_{p0,i}\}$ are $\{2,0.007,7\}$ and  $\{2,0.005,26\}$ for $i=3,4$ respectively. These combinations of parameters were chosen to provide large responses at different neighbourhoods of the environmental space, and hence to correspond to different frontier intervals. The distribution of maximum response $M_{R,100}$ for synthetic responses $R_3$, $R_4$ was estimated by generating multiple environmental simulations corresponding to periods of $100$ years, calculating response per sea state and storing only the maximum response observed and the values of $H_S$ and $T_P$ responsible for it. For responses $R_1$, $R_2$, \texttt{PPC} was used to extend the environmental model to include response; simulation under the model was then again used to accumulate the distribution of $M_{R,100}$. 

\begin{figure}
	\centering
	\includegraphics[width=0.9\textwidth]{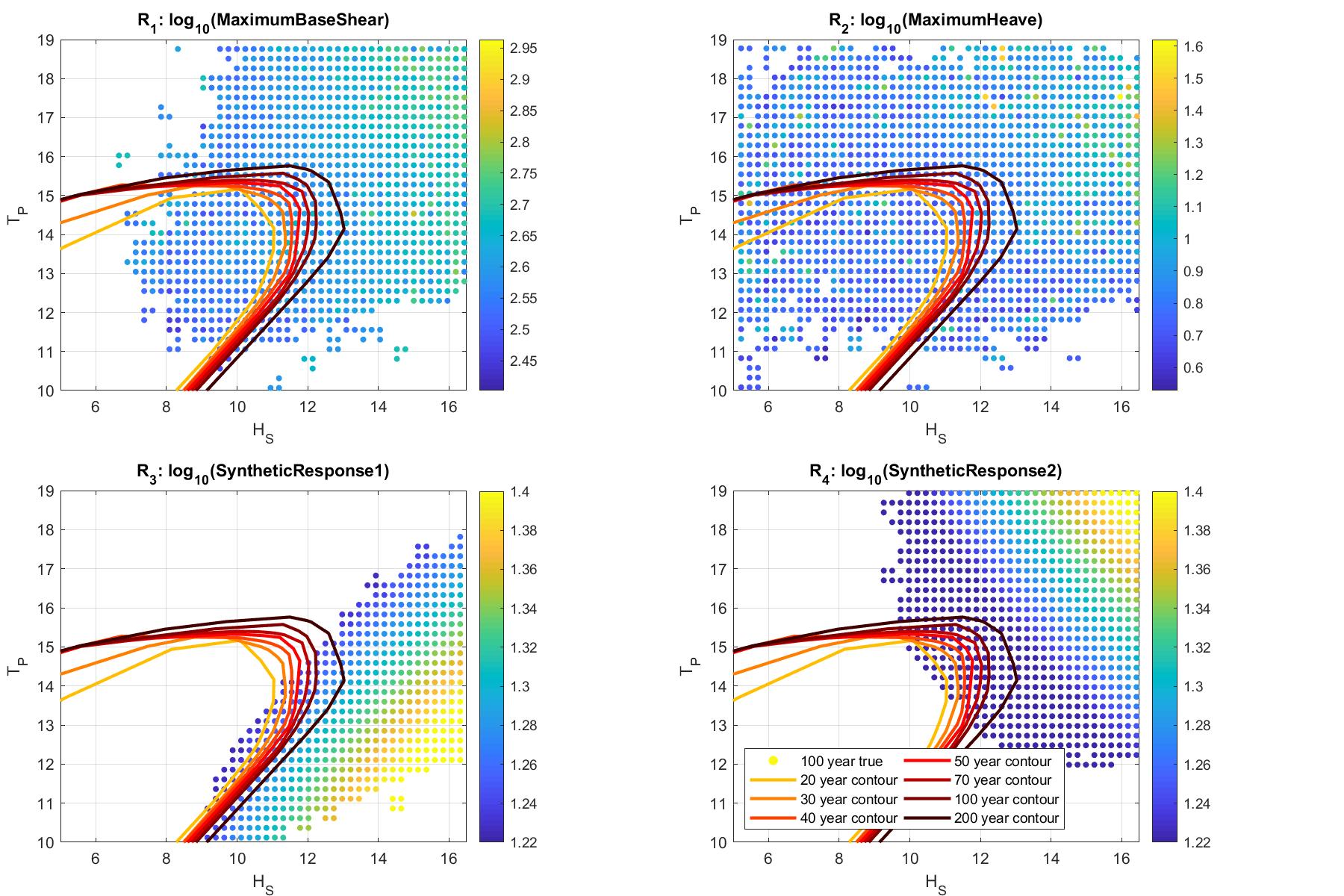}  
	\caption{Mean 100-year maximum responses $\{\bar{M}_{R_i,100}\}$ as a function of $H_S$ and $T_P$ estimated using 1000 realisations (of length 100 years) of $H_S$ and $T_P$. Points are coloured by the local mean value of maximum response estimated on a lattice of values for $H_S$ and $T_P$. Also shown are $N$-year $(H_S,T_P)$ direct sampling environmental contours for different values of $N$; contours are coloured yellow to dark brown by return period, in order of $N = \{20,30,40,50,70,100,200\}$ years. Panels on top row correspond to historic responses $R_1$ (left) and $R_2$ (right); panels on bottom row correspond to synthetic responses $R_3$ (left) and $R_4$ (right).}
	\label{Fgr:ToyScatter1}
\end{figure}

For each response in turn, the mean value $\bar{M}_{R,100}$ of the maximum 100-year response $M_{R,100}$ is plotted in Figure~\ref{Fgr:ToyScatter1}, and coloured by the value of $M_{R,100}$. Also plotted in the figure are direct sampling contours corresponding to 20, 30, 40, 50, 70, 100 and 200 years. Note that for each response $R_i$, only combinations of $H_S$ and $T_P$ giving rise to a least one occurrence of $M_{R_i,N}$ appear in the figure.
The figure shows typical features of the different responses. Synthetic response $R_3$ shows resonance effects $T_P \approx 13$s. Maximum base sheer ($R_1$) and synthetic response $R_4$ increase with increasing $H_S$ and $T_P$. This is true in general for maximum heave ($R_2$), but there are clearly large values of $M_{R_2,100}$ within even the 20-year environmental contour. That is, there are relatively benign environmental conditions, not even exceeding the $20$-year contour, which sometimes generate the $100$-year maximum response. For contours to be useful, we would expect to see the largest values of $100$-year maximum response lying outside the $100$-year contour, and smallest values of response within it. This is approximately the case for all responses, but certainly not always true for $R_2$.

\begin{figure}
	\centering
	\includegraphics[width=0.9\textwidth]{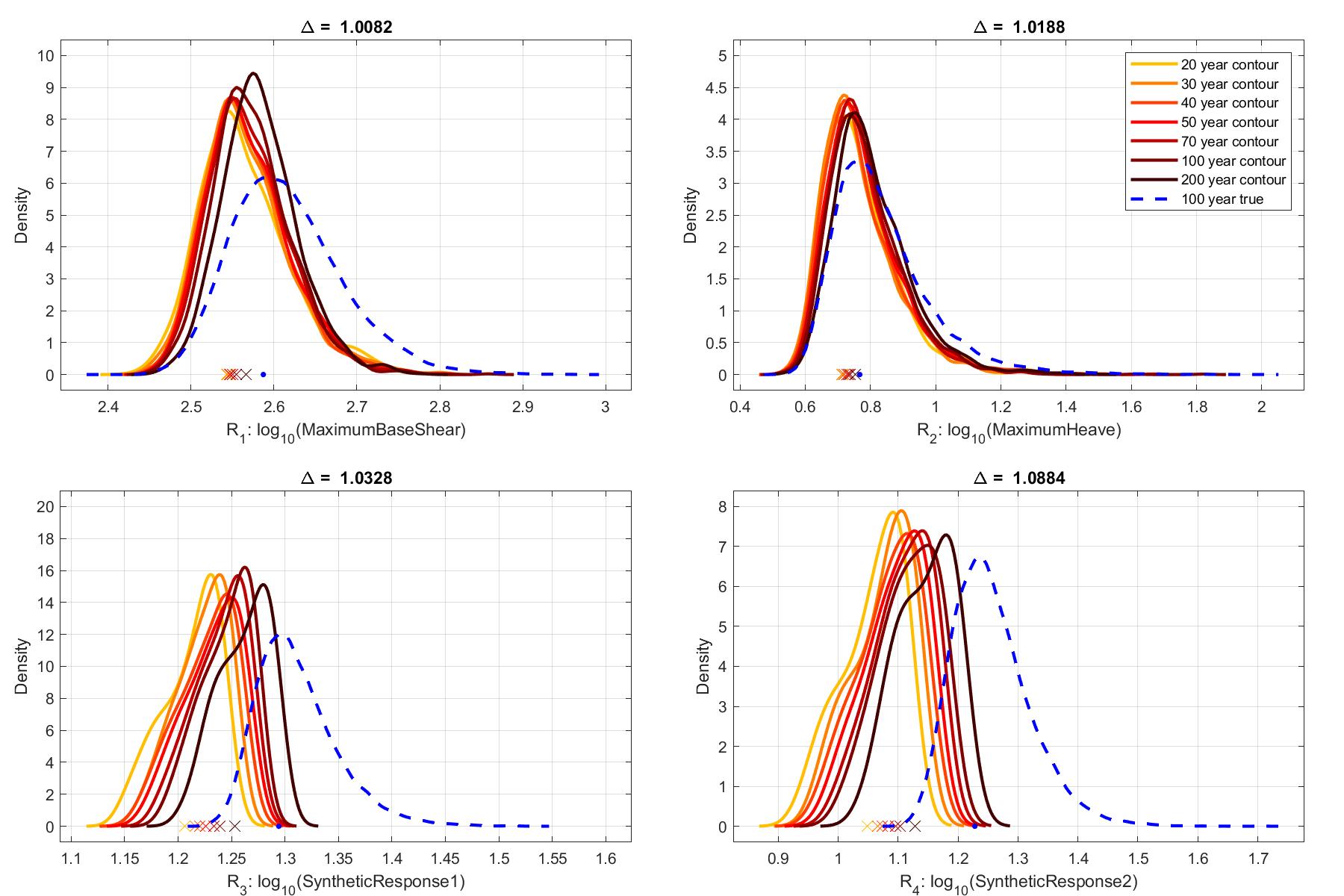}  
	\caption{Kernel density estimates for $100$-year maximum responses $\{M_{R_i,100}\}$. Estimates based on direct simulation of response are shown in dashed blue. Other density estimates (solid lines) are calculated from $(H_S, T_P)$ combinations lying near the corresponding $N$-year direct sampling contour shown in Figure \ref{Fgr:ToyScatter1}, for $N = \{20,30,40,50,70,100,200\}$. Coloured crosses indicate the location of the quantile of the response distribution with non-exceedance probability $\exp(-1)$ along each contour; the blue dot gives the corresponding $\exp(-1)$ ``true'' response from the blue curve. The factor $\Delta$ by which the $\exp(-1)$ response of the $100$-year environmental contour would need to be inflated to give the true $\exp(-1)$ $100$-year response is given in the title to each panel.}
	\label{Fgr:ToyDensAll}
\end{figure}

The extent to which the maximum response on the $100$-year environmental contour agrees with the actual distribution of $M_{R,N}$ from simulation can be assessed by comparing an estimate for the distribution of the true response against that evaluated for conditions on the contour, as illustrated in Figure \ref{Fgr:ToyDensAll}. It shows kernel density estimates for $\{M_{R_i,100}\}$ estimated by direct simulation (in dashed blue; which can be regarded as ``the truth''). The figure also shows corresponding kernel density estimates for $f_{M_{R,100}\text{Frontier}}$ of $M_{R,100}$ from combinations of $(H_S, T_P)$ lying on the contour frontier (scenario (b), shown in Figure~\ref{Fgr:ToyScatter1}), for a range of choices of $N$.

There is an obvious ordering of response density estimates with increasing return period, particularly for responses $R_3$ and $R_4$ as might be expected. Further, the location of densities estimated from different $N$-year environmental contours agrees to some degree with that of the true density of $M_{R,100}$. Moreover, for most cases the location of the quantile (see Section~\ref{Sct:EstCnt:Adj} with $p_C=p_R$) of the distribution of maximum response with non-exceedance probability $\exp(-1)$ (henceforth the ``$\exp(-1)$'' value) of the density estimate from the $100$-year environmental contour is in reasonable agreement with the location of the true $\exp(-1)$ value. Following Section~\ref{Sct:EstCnt:Adj}, defining $\Delta$ as the ratio of the $\exp(-1)$ quantile of the true $100$-year maximum response to the $\exp(-1)$ quantile of the distribution from the $100$-year environmental contour, we see that the environmental contour approach underestimates the $\exp(-1)$ response by between 1\% and 9\% for these examples. 

\begin{figure}
	\centering
	\includegraphics[width=0.9\textwidth]{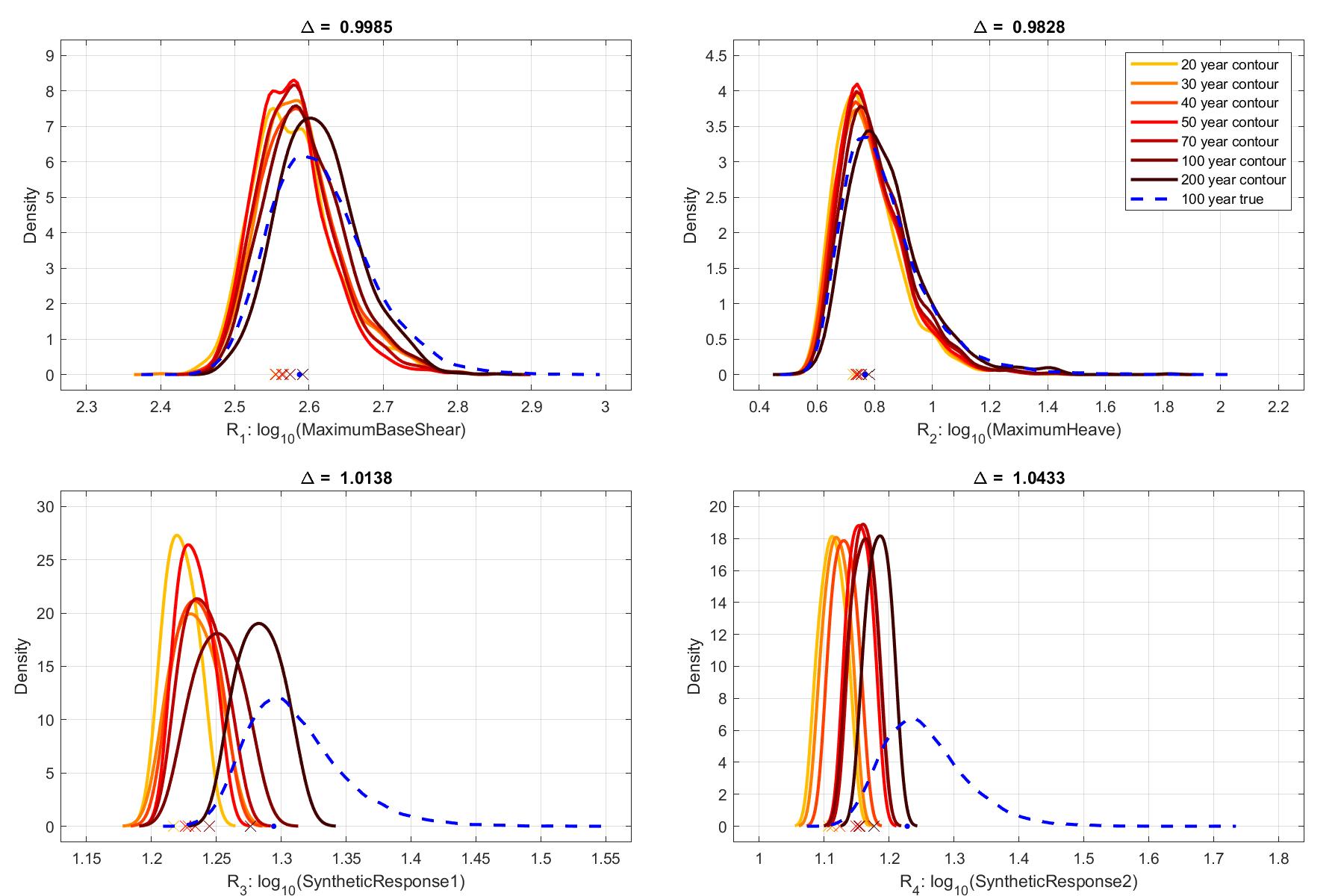}     
	\caption{Kernel density estimates for $100$-year maximum responses $\{M_{R_i,100}\}$. Estimates based on direct simulation of response are shown in dashed blue. Other density estimates (solid lines) are calculated from $(H_S, T_P)$ combinations near the point on the corresponding $N$-year direct sampling contour (Figure~\ref{Fgr:ToyScatter1}) \textit{corresponding to maximum $H_S$}, for $N = \{20,30,40,50,70,100,200\}$. Coloured crosses indicate the location of the quantile of the response distribution with non-exceedance probability $\exp(-1)$ along each contour; the blue dot gives the corresponding $\exp(-1)$ ``true'' response from the blue curve. The factor $\Delta$ by which the $\exp(-1)$ response of the $100$-year environmental contour would need to be inflated to give the true $\exp(-1)$ $100$-year response is given in the title to each panel.}
	\label{Fgr:ToyDensPnt}
\end{figure}

We next perform a similar comparison of response distributions, this time using only ($H_S$, $T_P$) combinations near the point on the contour with maximum $H_S$ (that is, scenario (a), to estimate $f_{M_{R,100}\text{Point}}$). Results, shown in Figure \ref{Fgr:ToyDensPnt}, have similar general characteristics to those of Figure~\ref{Fgr:ToyDensAll}. Values of $\Delta$ in the interval $(0.98,1.04)$ are estimated. In the current illustrations, therefore, it appears that both scenarios (a) and (b) provide reasonable estimates for the  $\exp(-1)$ quantile of $M_{R,100}$.

\subsection{Case study 2} \label{Sct:Prc:Cas2}
%
Here we extend the study of Section~\ref{Sct:Prc:Cas1} for responses $R_1$ (maximum base shear) and $R_2$ (maximum heave), specifically to make a comparison of direct sampling contours, joint exceedance contours (Section~\ref{Sct:EstCnt:JntExcCnt:JntEA}), and isodensity contours (from a conditional extremes analysis in Section~\ref{Sct:MdlEnv:Cnd}). For brevity, these approaches are henceforth referred to as ``direct sampling'', ``joint exceedance'' and ``empirical density'' respectively in this section.

\begin{figure}
	\centering
	\includegraphics[width=1\textwidth]{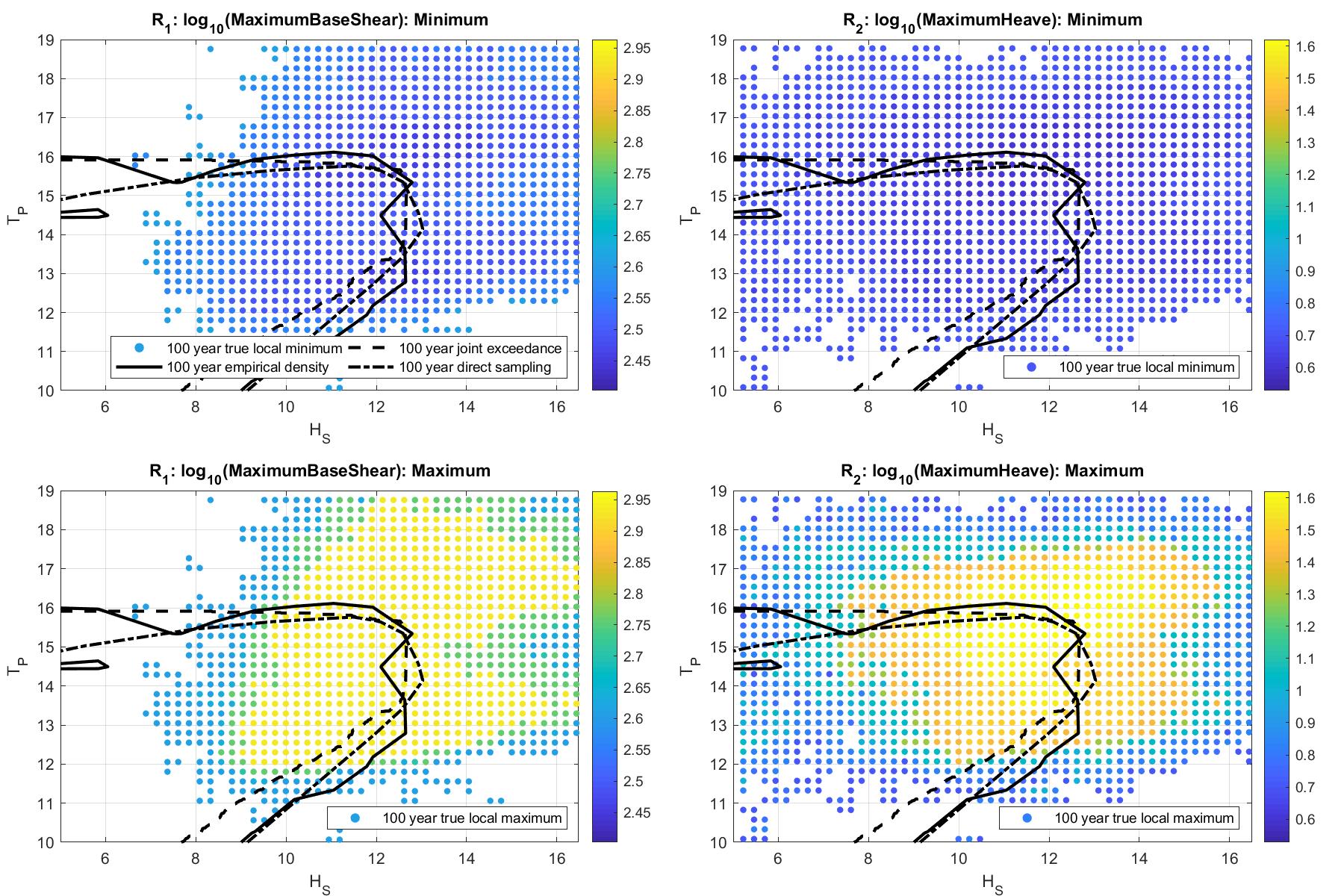}
	\caption{Minima ($\{M_{R_i,100}^{\text{min}}\}$, top row) and maxima ($\{M_{R_i,100}^{\text{max}}\}$, bottom row) of the 100-year maximum response, as a function of $H_S$ and $T_P$ estimated using 1000 realisations (of length 100 years) of $H_S$ and $T_P$ for responses $R_1$ (left) and $R_2$ (right). Points are coloured by the local minimum (top) or maximum (bottom) value of maximum response estimated on a lattice of values for $H_S$ and $T_P$. Each panel also shows $100$-year environmental contours from each of the direct sampling (black dot-dashed),  joint exceedance (black dashed) and CE density (black solid) methods.}
	\label{Fgr:Toy1}
\end{figure}

Figure~\ref{Fgr:Toy1} shows minima $\{M_{R_i,100}^{\text{min}}\}$ and maxima $\{M_{R_i,100}^{\text{max}}\}$ values of maximum responses $\{M_{R_i,100}\}$ from the same $1000$ simulations used to generate Figure~\ref{Fgr:ToyScatter1}. The colour of each disc in the top row indicates the value of the \textit{minimum} $100$-year maximum response seen for that combination of $H_S$ and $T_P$, using the same algorithm as for Figure~\ref{Fgr:ToyScatter1} to identify near neighbours. The bottom row shows corresponding values of \textit{maximum} $100$-year maximum response. It is clear that there is considerable variability in response for a given pair of values for $H_S$ and $T_P$. $100$-year environmental contours from each of the direct sampling, joint exceedance and empirical density methods are also shown in the figure. All contours have a similar frontier interval. There is good agreement between the direct sampling and joint exceedance contours in particular on the frontier interval; this is not surprising since the underlying methods have similar motivations. For the same simulation size, the empirical density contour is more difficult to estimate without applying considerable smoothing. Comparing the top and bottom rows of the figure, it also appears that the natural variability in the response (near the frontier interval of the contours) dominates any variability in the value of response \textit{along} the contours. In this case, therefore, none of the contours is particularly preferable; any of them would give approximately the same quality of estimate for $M_{R,100}$. 

It is interesting and intuitively appealing that the (yellow) area of largest values of maximum response (on the bottom row of the figure) is centred approximately on the frontier interval of the contours. However, for synthetic response $R_3$ in Figure \ref{Fgr:ToyScatter1}, we see that the frontier interval is offset (to lower $T_P$) from part of the environmental contour corresponding to largest $H_S$: focussing on an interval on the contour corresponding to largest $H_S$ to estimate $M_{R,100}$ would seem particularly suspect in this case, regardless of the choice of contour method.

Overall, it appears that the key to success is ensuring that the response is quantified (using time-domain simulation or other) at a sufficient number of ($H_S$, $T_P$) combinations on or near the frontier interval of any reasonably well-defined contour.

\section{Discussion and conclusions} \label{Sct:DscCnc}
%
As shown in Section~\ref{Sct:EstCnt}, environmental contours provide useful characterisations of the extent of the joint distribution of environmental variables. Some contour methods assume particular parametric forms for the (conditional) distributions of environmental variables; other methods generate convex contours on particular scales; other contour approaches are only defined on part of the domain of environmental variables. There is concern in the user community that a contour should ``look right'', closely hugging the boundaries of scatter plots of historic or simulated environmental variables.

The usual motivation for applying a contour approach in ocean engineering is to find environmental conditions efficiently (for a return period of $N$ years say) which will generate approximately the $N$-year maximum response. Environmental contours therefore provide a means of reducing the burden of running full long-term response analysis for a wide range of environmental conditions. Different types of environmental contours find favour based on their ability to estimate the $N$-year maximum response from the $N$-year environmental contour. 

An environmental contour is estimated with no regard whatsoever to structural details. Since environmental contours are independent of structural specifics, they can then be used in principle to study different structures in a given environment provided that the underlying assumptions linking environment and structure are not violated. 

There is no fundamental link between points on an environmental contour and structural response in general, and no reasonable expectation therefore that points on the $N$-year environmental contour should yield the $N$-year maximum response. Attempts have been made to compare different environmental contour methods using a response-based criterion, although it is mathematically obvious and generally recognised in the user community that no single approach is appropriate for all structure and response types, and that considerable ambiguity will always remain. The manner in which an environmental contour relates to extreme response depends on the specifics of the structure. However, for typical $H_S$-driven structures, empirical evidence suggests the responses generated from points along the environmental contour in ($H_S$, $T_P$)-space for a given return period are reasonable estimates of the actual maximum response corresponding to the same return period. In the presence of resonant response and non-extreme values of $T_P$, using points from the contour near the maximum $H_S$ can be misleading, since the response is not completely $H_S$-dominated. It is critical therefore that the dominant environmental variables are included in the estimation of environmental contour. It is apparent from physical considerations that extreme occurrences of some structural responses should not coincide with those of extreme environmental variables; the $N$-year environmental contour is unlikely to provide any guidance regarding the $N$-year maximum response for such responses. We also note methods to adjust (or inflate) contours to correct for sources of bias (for estimation of extreme response) including the effects short-term variability, violation of (marginal and dependence) modelling assumptions, uncertainty in parameter estimates, etc.

There is some debate within the user community regarding the relative merits of using observations of the environmental variables for serially-dependent sea states, compared with near-independent storm peak characteristics. Given that the rates of occurrence of events are taken into account, both can provide useful estimates of joint models for the environment and hence environmental contours. The advantage of using sea state data is that sample size is large, potentially allowing a more detailed description of the joint distribution of environmental parameters to be estimated. However, because sea state data is serially-correlated, naive estimates of uncertainties for model parameters and inferences under the model will be too small, but can be corrected (for example, using sandwich estimators or bootstrap resampling).

Multi-modal distributions of environmental variables can be caused by different physical processes or by covariate effects (for example, fetch length as a function of direction); in these cases, isodensity contours may be more reasonable summaries of the joint distribution than those based on joint exceedance, since there may be regions of low probability between modes. Further it may be useful to partition the environmental space by covariate and perform separate analyses per partition. Alternatively, it is also possible to estimate joint models incorporating covariate effects. 

Basing design conditions on the $N$-year environmental contour alone neglects sources of bias and variability in estimation of the $N$-year maximum response, including short-term variability in response. Approximate methods to inflate the environmental contour, or adjust its return period, are available. It is always possible, given fully-specified environments and structural responses, to estimate inflation factors which map some quantile of the distribution of $N$-year maximum response onto the corresponding quantile of the distribution of maximum response given sea states on the environmental contour. However, the value of inflation factor in general will be a function of quantile level, the structure and the response. It is likely that estimating inflation factors (or adjusting contour return periods) based on comparing central characteristics (for example, mean, median or mode) of response distributions will prove more stable, since Monte Carlo simulations of a given size provide better estimates of central characteristics than those of tails.

In some applications, it may be that environmental contours will be used to estimate multiple correlated responses. In such cases, care needs to be taken that the contour is used to estimate the responses \textit{jointly} corresponding to a given return period, rather than estimating independent marginal return values.

As can be seen from the end-user survey in \ref{SctApp:SrvFnd}, there are valid concerns regarding a sensible definition of an environmental contour, its estimation and adjustment in relation to structural response. There are many articles in the literature comparing different contour methods, inflation factors etc. However, in reality, most sensibly-proposed contour methods (including the direct sampling, joint exceedance and empirical density examined here) locate the frontier interval of the contour within the same region of environmental space. Uncertainties due to the details of the structure, and even when the structure is defined to the natural variability of structural response given the environment, are in general issues of far greater concern. Any reasonable choice of contour, given the considerations explored in this paper, will suffice.

Methods such as IFORM and the direct sampling method are advantageous in that they impose a link between environment and structure by make assumptions about the characteristics of failure surfaces as a function of the environmental variables. Given these assumptions, it is possible to link the exceedance probability associated with a given environmental contour with structural failure probability. Although conditions from an $N$-year environmental contour need not result exactly in $N$-year responses, IFORM and direct sampling provide at least some understanding of how an $N$-year environmental contour is related to the $N$-year maximum response. Both IFORM and direct sampling approaches assume a linearised failure boundary. The basic difference between the approaches arises from the fact that linearisation for IFORM is performed in the transformed $\un{U}$-space, and in direct sampling approach in the original $\un{X}$-space of environmental variables (e.g. \citealt{VnmBtn15, Vnm17} and references therein). For both IFORM and direct sampling contours, the relationship established is between the exceedance probability associated with the contour (on some scale) and the probability of structural failure. This does not guarantee however that searching along an IFORM or direct sampling contour for return period $T$ will isolate the key features of the $T$-year maximum response; the relative performance of IFORM and direct sampling in estimating extreme responses is application-dependent. 

For better specification of design conditions, a response model is necessary. To determine the frontier interval of a design contour within the domain of environmental variables, inflation factors for contours etc., some knowledge of, or working assumption regarding the response is required. As specification of the response evolves, sufficient, for example, to estimate the frontier interval and inflation factors, arguably there is already sufficient information (or supposition) to use better models to describe approximate responses and their uncertainties. From a statistical perspective, it is hard to avoid the impression that, in terms of estimating extreme structural response, an environmental contour is just an approximation to a sample from the tail of the distribution of environmental variables. In fact, for most purposes, an appropriate sample from the tail of the distribution of environmental variables would be preferable. Furthermore, with the advent of methods such as statistical emulation (used widely in approximating complex physical systems, including metocean design e.g. \citealt{GldEA14, JnsEA18}), a computationally-efficient approximate response model can be estimated, along with its uncertainty, for many if not all applications. Given this, the emulator would provide a mechanism for response-based design in all situations, avoiding the need for environmental contours completely. It is apparent that the coastal engineering community is already moving in this direction (e.g. \citealt{GldEA17}); a key assumption with statistical emulation is that a representative set of cases is available to quantify all important relationships between environment and structural loading. Nevertheless, part of the user-cited strength of the environmental contour method is its relative simplicity and computational efficiency. There is clearly a trade-off between a thorough probabilistic response-driven analysis (which may be technically more involved and less familiar to the practitioner) compared to a simple approximate approach (with which the user is relatively familiar and confident). 

Given the above considerations, and comments in the end-user survey, to assist the practitioner in deciding when and where to use an environmental contour approach, we present the following brief check-list. We recommend the use of environmental contours in the following circumstances.

\subsection{When to use environmental contours} \label{Sct:DscCnc:When}
%
\pbi
\item \textbf{Nature of responses and environmental variables are known}: The dominant structural responses are all known. The dominant environmental variables driving each structural response are all known, and the value of response is dominated by long-term variability of the environmental variables: extreme environments produce extreme responses. The influence of short-term environmental variability is relatively small.  [There are some responses (for example, fatigue) that are not dominated by extremes of the environment; environmental contour methods are therefore not appropriate.] 
\item \textbf{Response-based analysis is not possible}: (a) There are no adequate computationally-efficient structural response models available. [If these models are available, a response-based analysis should be performed.] (b) There are computationally-demanding structural response models available, but no time or expertise to develop approximate structural response models (for example, generic load models, statistical emulators) using these. [If these models can be estimated, a response-based analysis should be performed.]
\item \textbf{At outline design stage}: The specifics of the structure may not be clear at outline design. For this reason, the environmental contour may provide a useful source of extreme sea states suitable for evaluating a range of different structures.

\pei

Once it has been decided that an environmental contour approach is suitable, the following are then recommended.

\subsection{How to use environmental contours wisely} \label{Sct:DscCnc:How}
\pbi
\item \textbf{Reality check}: Remember that environmental contours are approximate methods that can only provide approximations to extreme responses. The use of contour approaches may need to be supported in final design by full long-term analysis.
\item \textbf{Sufficient environmental data available}: There are sufficient historical data available to estimate the joint distribution of all these environmental variables adequately using, for example, a method from Section~\ref{Sct:MdlEnv}.
\item \textbf{Estimate more than one environmental model, and consider the sensitivity of the model to arbitrary modelling choices}: The sensitivity of environmental contour estimates to arbitrary choices made when estimating a model for the joint distribution of environmental parameters should be investigated. [When different equally-plausible environmental models provide different contour estimates, the current research suggests that all contours should be considered valid and used together for choice of environmental values corresponding to extreme responses. Equally, the user might well be concerned when two different environmental models provide materially different contour estimates (using a common contouring approach).]
\item \textbf{If unsure which contour to use, estimate more than one type}: Each type of environmental contour is seeking to achieve different objectives. If you are not clear which contour is most suitable for your application, consider estimating contours of different types, and establish approximate consistency of inferences from different contours. [When different equally-plausible contours give materially different results, all contours should be considered valid and used together for choice of environmental values corresponding to extreme responses.]
\item \textbf{Choose multiple points from the environmental contours for response evaluation}: Multiple combinations of values of environmental variables falling on or near the frontier interval of the environmental contour should be used. [When the frontier interval is not known, a wide set of combinations of values of environmental variables on or near the environmental contour should be used. If in doubt, choose more points and choose points more widely.]
\item \textbf{Consider other sources of uncertainty}: (a) How influential are the effects of covariates (directionality, seasonality) [If important, fit a non-stationary environmental model (for example, using PPC from Section~\ref{Sct:Prc})];  (b) Have all environmental variables influencing the response been considered in the environmental model and contours? [If not, consider estimating environmental models and contours in higher dimensions].
\pei

\section{Acknowledgement} \label{Sct:Ack}
%
This work was part-funded by the European Union ERA-NET project entitled ``Environmental Contours for SAfe DEsign of Ships and other marine structures (ECSADES)''. The \texttt{PPC} software and user guide is freely-available as MATLAB code from the authors. We thanks Kevin Ewans for useful comments on the manuscript.

\processdelayedfloats

\clearpage
\appendix
\section{Supporting information} \label{SctApp:SppInf}
\renewcommand\thefigure{A.\arabic{figure}}    
\renewcommand\thepostfigure{A.\arabic{postfigure}} 
\setcounter{figure}{0}  
\setcounter{postfigure}{0}  

\subsection{Survey findings} \label{SctApp:SrvFnd}
%
Further details of the survey summarised in Section~\ref{Sct:Int:Srv} are given here. For brevity, the respondents full answers are summarised here. The full (anonymised) survey responses can however be provided on request.

The survey questions were distributed to industry/consulting contacts of those involved in the ECSADES project as well as authors of key literature in the area. We do not claim that this sample is representative of the user community for contours, but hopefully it is informative. Of the 19 respondents, a large number were based in Norway with contributions also from the US, the Netherlands, the UK and Australia; with 7 from academia and 12 from industry/consulting. 

\begin{enumerate}
	\item How many times have you used environmental contours as part of your work in the last 3 years?
	\begin{itemize}
		\item Different frequencies of use: from daily to hardly ever
		\item Geographic variability: frequent use cited by respondents based in Norway, less frequent in the U.S.
		\item Contours form an integral part of reliability assessment / design practices
		\item Academically, contours garner less interest. Appearing mainly in post-graduate projects with industrial applications.
	\end{itemize}
	
	\item What kind of environmental contours do you use (for example, derived from FORM, or the conditional extremes model of Heffernan \& Tawn, or other)?
	\begin{itemize}
		\item Large user group for FORM/IFORM 
		\item Conditional extremes approach of \cite{HffTwn04} 
		\item Marginal and dependence modelling based on threshold exceedance distributions
		\item AND and JOIN-SEA cited for coastal applications
		\item Kernel density estimation
		\item The need for contours in any form also questioned in the event that direct response-based analysis is possible
	\end{itemize}
	
	\item For what purpose do you use environmental contours? Can you describe as precisely as possible the information you take from the environmental contour, and how you use this? Do you use the whole contour or just part of it? Which part of the contour, and why? Do you actually need information about the whole contour? Do you take any steps to account for uncertainties and potential biases?
	\begin{itemize}
		\item Two dimensional contour representing ``N-year'' environmental conditions is typically estimated (never higher dimensions)
		\item Response analysis performed on a subset of points along the environmental contour	
		\item Region of the contour over which this subset of points is focussed can be motivated by physical understanding of mechanisms driving response, for example, wave-dominated or resonant response
		\item Sometimes only ``short term'' response analysis performed; additional extrapolation from ``short-term'' to ``long-term'' necessary.
		\item N-year structural response is estimated from the responses resulting from conditions defined by points on the environmental contour by:
		\begin{itemize}
			\item Inflating the response somehow: 90\% quantile, median $\times$ 1.3 or other ad-hoc method, to adjust for short-term variability
			\item Inflate “environment”: that is, explore responses for a contour representing a longer return period (N)
		\end{itemize}
		\item Bias, (epistemic) uncertainty and inherent (aleatory) randomness is almost never incorporated
	\end{itemize}
	
	\item What do you think are the advantages of using environmental contours?
	\begin{itemize}
		\item Environmental contours provide a reasonable approximate approach to estimate N-year responses
		\item Industry-accepted
		\item Calculated without any knowledge of structure being designed; independent of the response; same contour can be used for range of responses
		\item Simple to use, inexpensive, quick - especially for typical cases (for example, Hs-Tp)
		\item Computational efficiency compared full time-domain analysis
	\end{itemize}

	\item What, in your opinion, are the disadvantages or problems associated with using environmental contours? Are there specific circumstances in which you find that environmental contours do not work well or are difficult to use?
	\begin{itemize}
		\item Concern about details of choosing response from a subset estimated along the contour: using arbitrary quantiles and scale factors etc.
		\item Simulations using short-term responses used to estimate long-term response; high quantile necessary.
		\item Environmental contour method does not involve the response directly (so arbitrary calibration of response will always be necessary); “response-based” analysis is a better approach when possible
		\item Clients do not understand what the contour represents and how it has been calculated, can therefore sometimes be a  “hard sell”
		\item Lots of approaches to defining the contour, not clear which (if any) is better
		\item Naïve application of statistical methods can yield physically unreasonable contours (for example, non-physical wave steepness)
		\item Approaches to extending to higher dimensions (3D and above) or to multiple responses are more ad-hoc, incorporating correct physics even more problematic
		\item Need to check conclusions using full-scale simulation / long-term analysis
		\item No obvious way to estimate the \textit{distributional} properties of N-year response
		\item Uncertainties are rarely quantified
	\end{itemize}
	
	\item Do you use design guidelines or other literature sources to guide your use of environmental contours? Could you list these?
	\begin{itemize}
		\item  Statoil procedures
		\item  Norsok N003
		\item  DNV-RP-C205 (2014)
		\item  FORM/iFORM: \cite{WntEA93}
		\item  JOIN-SEA: \cite{HksEA02}
		\item  \cite{HffTwn04}
		\item  Shell LSM model for response-based design
	\end{itemize}
\end{enumerate}

\subsection{\texttt{PPC} model}  \label{SctApp:PPC}
%
The penalised piecewise constant (\texttt{PPC}) extreme value model allows the estimation of non-stationary marginal and conditional extremes for peaks over threshold using a simple description of non-stationarity with respect to covariates in marginal and dependence models. An early deployment of the \texttt{PPC} model as software is described in \cite{RssEA17b}; the current version of the \texttt{PPC} software, developed as part of the ECSADES project (see Section~\ref{Sct:Ack}), is freely available from the authors. For each observation $\un{x}_i$ in the sample $\{\un{x}_i\}$, we assume that an associated (potentially vector) covariate $\un{\theta}_i$ is available. The value of covariate $\un{\theta}_i$ is used to allocate the observation to one and only one of $n_C$ covariate intervals $\{C_k\}_{k=1}^{n_C}$ by means of an allocation vector $A$ such that $k=A(i)$ and $\{\un{x}_i\}=\bigcup C_k$. For each $k$, all observations in the set $\{\un{x_{i'}}\}_{A(i')=k}$ with the same covariate interval $C_k$ are assumed to have common extreme marginal and dependence characteristics. 

Non-stationary marginal extreme value characteristics of each variate $X_j$ are then estimated in turn using a generalised Pareto model and cross-validated roughness-penalised maximum likelihood estimation. For variable $X_j$ and covariate interval $C_k$, the extreme value threshold $\psi_{jk}>0$ is assumed to be a quantile of the empirical distribution of the variate in that interval, with specified non-exceedance probability $\tau_j \in (0,1)$, with $\tau_j$ constant across intervals, and estimated by counting. Threshold exceedances are assumed to follow the generalised Pareto distribution with shape $\xi_{j} \in \mathbb{R}$ and scale $\sigma_{jk}>0$. $\xi_{j}$ is assumed constant (but unknown) across covariate intervals, and the reasonableness of the assumption assessed by inspection of diagnostic plots. Parameters $\xi_j,\{\sigma_{jk}\}$ are estimated by maximising the predictive performance of a roughness-penalised model, optimally regulating the extent to which $\{\sigma_{jk}\}$ varies across interval, using a cross-validation procedure.

After marginal fitting, the sample $\{\un{x}_i\}$ is transformed to standard Laplace scale as $\{\un{\tilde{x}}_i\}$ and the conditional extremes model outlined in Section~\ref{Sct:MdlEnv:Cnd} fitted. Following the notation of that section, for each choice of conditioning variate $X_q$, linear parameters $\un{\alpha}_{-q}$ for the conditioned variates vary across covariate bins with variation regularised using cross-validation to optimise predictive performance for each response in turn. The corresponding value of exponent parameters $\un{\beta}_{-k}$ is assumed constant with respect to covariates. Sets of residuals $\mathcal{R}_{-q}$ from the fit, for each choice of $q$ are also isolated.

Simulation under the fitted model, and transformation of realisations to their original marginal scale is then possible, corresponding to a return period of arbitrary length. In particular, samples simulated under the model can be used to estimate environmental contours of different types, and explore their relative characteristics. In practice, the full \texttt{PPC} modelling procedure is repeated for $n_U$ bootstrap resamples of the original sample to capture sampling uncertainty, each resample using a different choice of marginal and dependence thresholds to capture threshold uncertainty. Estimates for marginal and dependence parameters therefore correspond to $n_U$ arrays of values capturing sampling and threshold specification uncertainty, which are used to propagate uncertainty from these sources into simulations under the model.

We use the \texttt{PPC} model to estimate a number of the environmental contours discussed in Section~\ref{Sct:EstCnt} and investigate their characteristics, in particular their relationship to extremes of structural response. Specifically, because of its recently popularity, we consider the direct sampling contour (Section~\ref{Sct:EstCnt:JntExcCnt:DrcSmpCnt}); we also consider the joint exceedance contour outlined in Section~\ref{Sct:EstCnt:JntExcCnt:JntEA}.  Both these contour methods use a simulated sample under the environmental model as starting point. Further we consider an isodensity contour (Section~\ref{Sct:EstCnt:IsoDnsCnt} based on the conditional extremes model of Section~\ref{Sct:MdlEnv:Cnd}; this approach infers the contour directly from the properties of the environmental model with no need for simulation.

\processdelayedfloats

\pagebreak
\section*{References}
\bibliographystyle{elsarticle-harv}
\bibliography{phil}

\end{document}